\documentstyle[12pt]{article}
\newtheorem{theorem}{Theorem}

\begin{document}

\title{Locally Anisotropic Structures and Nonlinear\\
 Connections in Einstein and Gauge Gravity }
 \author{ Sergiu I.\ Vacaru  \thanks{ e--mail:\
vacaru@lises.asm.md, on leave of absence from
 the Institute of Applied Physics,\newline Academy of Sciences,
  Chi\c sin\v au MD2028, Republic of Moldova } \quad
   and Heinz Dehnen\thanks{e--mail:\
 Heinz.Dehnen@uni-konstanz.de} \\[6pt]\\
\small  Fachbereich Physik, Universitat Konstanz, \\ \small
Postfach M 638, D--78457, Konstanz, Germany
   }
\date{August 15, 2000}
\maketitle

\abstract We analyze local anisotropies induced by anholonomic
frames and associated nonlinear connections in general relativity
and extensions to affine--Poincar\'e and de Sitter  gauge gravity
and different types of Kaluza--Klein theories.  We construct some
new classes of cosmological solutions of gravitational field
 equations describing Friedmann--Robertson--Walker like universes with
 rotation (ellongated and flattened) ellipsoidal or torus symmetry.

\endabstract


\section{Introduction}

The search for exact solutions with generic local anisotropy in
general relativity, gauge gravity and non--Riemannian extensions
has its motivation from low energy limits in modern string and
Kaluza--Klein theories. Such classes of solutions constructed by
using moving anholonomic frame fields (tetrads, or vierbeins; we
shall use the term frames for higher dimensions) reflect a new
type of constrained dynamics and locally anisotropic interactions
of gravitational and matter fields \cite{v1,v2}.

What are the requirements of such constructions and their physical
treatment? We belive that such solutuions should have the
properties:\ (i) they satisfy the Einstein equations in general
relativity and are locally anisotropic generalizations of some
known solutions in isotropic limits with a well posed Cauchy
problem;\ (ii) the corresponding geometrical and physical values
are defined, as a rule, with respect to an anholonomic system of
reference which reflects the imposed constraints and supposed
symmetry of locally anisotropic interactions; the reformulation of
results for a coordinate frame is also possible;\ (iii) by
applying the method of moving frames of reference, we can
generalize the solutions to some analogous in metric--affine
and/or gauge gravity, in higher dimension and string theories.

Comparing with the previos results \cite{vjmp,vnp,vjhep,vhp,vg} on
definition of self--consistent field theories incorporating
various possible anisotropic, inhomogeneous and sto\-chas\-tic
manifestations of classical and quantum interactions on locally
anisotropic and higher order anisotropic spaces, we emphasize
that, in this paper, we shall be interested not in some extensions
of the well known gravity theories with locally isotropic
spacetimes ((pseudo)Riemannian or Riemanian--Cartan--Weyl ones, in
brief, RCW--spacetimes) to Finsler geometry and its
generalizations, but in a proof that locally anisotropic
structures (Finsler, Lagrange and higher order developments
\cite{finsler,car35,run,matsumoto,asanov,ma94,m97,bejancu,gb})
could be induced by anholonomic frames on locally isotropic
spaces, even in general relativity and its metric--affine and
gauge like modifications \cite
{hgmn,tseyt,p,pd,mielke,deh,vg,pbo,wal}.

To evolve some new (frame anholonomy) features of locally
isotropic gravity theories we shall apply the methods of the
geometry of anholonomic frames and associated nonlinear connection
(in brief, N--connection) structures elaborated in details for
bundle spaces and generalized Finsler spaces in monographs
\cite{ma94,m97,bejancu} with further developments for spinor
differential geometry, superspaces and  stochastic calculus in
\cite {vjmp,vnp,vjhep,vhp}. The first rigurous global definition
of N--connections is due to W. Barthel \cite{barth} but the idea
and some rough constructions could be found in the E. Cartan's
works \cite{car35}. We note that the point of this paper is to
emphasize the generic locally anisotropic geometry and physics and
apply the N--connection method for ´non--Finslerian´ (pseudo)
Riemannian and RCW spacetimes. Here, it should be mentioned that
anholonomic frames are considered in detail, for instance, in
monographs \cite{eh,mtw,pr} and with respect to geometrization of
gauge theories in \cite{mielke,pbo} but not concerning the topic
on associated N--connection structures which grounds our geometric
approach to anisotropies in physical theories and developing of a
new method of integrating gravitational field equations.

The paper is organized as follows:\ Section 2 contains a brief
introduction into the geometry of anholonomic frames and
associated nonlinear connection structures in (pseudo) Riemannian
spaces. Section 3 is devoted to the higher order anisotropic
structures in Einstein gravity. In Section 4 we formulate the
theory of gauge (Yang-Mills) fields on higher order anisotropic
spaces; the variational proof of gauge field equations is
considered in connection with a ''pure'' geometrical method of
definition of field equations. In Section 5 the higher order
anisotropic gravity is reformulated as a gauge theory for
nonsemisimple groups. A model of nonlinear de Sitter gauge gravity
with higher order anisotropy is formulated in Section 6. An ansatz
for generation of four dimenisonal solutions with generic
anisotropy of the Einstein equations is analyzed in Section 7.
Some classes of solutions, with generic anisotropy, of Einstein
equations describing Friedmann--Robertson--Walker like universes
with rotation
 (ellongated and flattened) ellipsoid and torus symmetry are
constructed in Section 8.  Concluding remarks are given in Section
9.

\section{ Anholonomic Frames on (Pseudo) Riemannian Spaces}
For definiteness, we consider a $\left( n+m\right) $--dimensional
(pseudo) Riemannian spacetime $V^{(n+m)},$ being a paracompact and
connected
Hausdorff $C^\infty $--manifold, enabled with a nonsigular metric%
$$ ds^2=\widetilde{g}_{\alpha \beta }\ du^\alpha \otimes du^\beta
$$ with the coefficients
\begin{equation}
\label{ansatz}\widetilde{g}_{\alpha \beta }=\left[
\begin{array}{cc}
g_{ij}+N_i^aN_j^bh_{ab} & N_j^eh_{ae} \\ N_i^eh_{be} & h_{ab}
\end{array}
\right]
\end{equation}
parametrized with respect to a local coordinate basis $du^\alpha
=\left( dx^i,dy^a\right) ,$ having its dual $\partial /u^\alpha
=\left(
\partial /x^i,\partial /y^a\right) ,$ where the indices of geometrical
objects and local coordinate $u^\alpha =\left( x^k,y^a\right) $
run correspondingly the values:\ (for Greek indices)$\alpha ,\beta
,\ldots =n+m;$ for (Latin indices) $i,j,k,...=1,2,...,n$ and
$a,b,c,...=1,2,...,m$ . We shall use 'tilds' if would be necessary
to emphasize that a value is defined with respect to a coordinate
basis.

The metric (\ref{ansatz}) can be rewritten in a block $(n\times n)
+ (m \times m) $ form
\begin{equation}
\label{dm}g_{\alpha \beta }=\left(
\begin{array}{cc}
g_{ij}(x^k,y^a) & 0 \\ 0 & h_{ab}(x^k,y^a)
\end{array}
\right)
\end{equation}
with respect to a subclass of $n+m$ anholonomic frame basis (for
four dimensions one used terms tetrads, or vierbiends) defined
\begin{equation}
\label{dder}\delta _\alpha =(\delta _i,\partial _a)=\frac \delta
{\partial u^\alpha }=\left( \delta _i=\frac \delta {\partial
x^i}=\frac \partial {\partial x^i}-N_i^b\left( x^j,y^c\right)
\frac \partial {\partial y^b},\partial _a=\frac \partial {\partial
y^a}\right)
\end{equation}
and
\begin{equation}
\label{ddif}\delta ^\beta =\left( d^i,\delta ^a\right) =\delta
u^\beta =\left( d^i=dx^i,\delta ^a=\delta y^a=dy^a+N_k^a\left(
x^j,y^b\right) dx^k\right) ,
\end{equation}
called the locally anisotropic bases (in brief, la--bases) adapted
to the coefficients $N_j^a.$ The $n\times n$ matrice $g_{ij}$
defines the so--called horizontal metric (in brief, h--metric) and
the $m\times m$ matrice $h_{ab}$ defines the vertical (v--metric)
with respect to the associated nonlinear connection
(N--connection) structure given by its coefficients $N_j^a\left(
u^\alpha \right) $ from (\ref{dder}) and (\ref{ddif}). The
geometry of N--connections is studied in detail in
\cite{barth,ma94};\ here we shall consider its applications with
respect to anholonomic frames in general relativity and its
locally isotropic generalizations.

A frame structure $\delta _\alpha $ (\ref{dder}) on $V^{(n+m)}$ is
characterized by its anholonomy relations
\begin{equation}
\label{anholon}\delta _\alpha \delta _\beta -\delta _\beta \delta
_\alpha =w_{~\alpha \beta }^\gamma \delta _\gamma .
\end{equation}
with anholonomy coefficients $w_{~\beta \gamma }^\alpha .$The
elongation of partial derivatives (by N--coefficients) in the
locally adapted partial derivatives (\ref{dder}) reflects the fact
that on the (pseudo) Riemannian spacetime $V^{(n+m)}$ it is
modelled a generic local anisotropy characterized by the
anholonomy relations (\ref{anholon}) when the anholonomy
coefficients are computed as follows
\begin{eqnarray}
w_{~ij}^k & = & 0,w_{~aj}^k=0,w_{~ia}^k=0,w_{~ab}^k=0,w_{~ab}^c=0,
\nonumber\\ w_{~ij}^a & = & -\Omega _{ij}^a,w_{~aj}^b=-\partial
_aN_i^b,w_{~ia}^b=\partial _aN_i^b, \nonumber
\end{eqnarray}
where%
$$ \Omega _{ij}^a=\partial _iN_j^a-\partial _jN_i^a+N_i^b\partial
_bN_j^a-N_j^b\partial _bN_i^a $$ defines the coefficients of the
N--connection curvature, in brief, N--curvature. On (pseudo)
Riemannian spacetimes this is a characteristic of a chosen
anholonomic system of reference.

A N--connection $N$ defines a global decomposition, $$
N:V^{(n+m)}=H^{(n)}\oplus V^{(m)}, $$
of spacetime $V^{(n+m)}$ into a $n$--dimensional horizontal subspace $%
H^{(n)} $ (with holonomic $x$--coordinates) and into a
$m$--dimensional
vertical subspace $V^{(m)}$ (with anisotropic, anholonomic, $y$%
--coordinates). This form of parametrizations of sets of mixt
holonomic--anholonomic frames is very useful for investigation,
for instance, of kinetic and thermodynamic systems in general
relativity, spinor and gauge field interactions in curved
spacetimes and for definition of non--trivial reductions from
higher dimension to lower dimension ones in Kaluza--Klein
theories. In the last case the N--connection could be treated as a
'splitting' field into base's and extra dimensions with the
anholonomic (equivalently, anisotropic) structure defined from
some prescribed types of symmetries and constraints (imposed on a
physical system) or, for a different class of theories, with some
dynamical field equations following in the low energy limit of
string theories \cite{vap,vnp} or from Einstein equations on a
higher dimension space.

The locally anisotropic spacetimes, la--spacetimes, to be
investigated in this section
 are considered to be some
(pseudo) Riemannian manifolds $V^{(n+m)}$ enabled with a frame, in
general, anholonomic structures of basis vector fields, $\delta
^\alpha =(\delta ^i,\delta ^a)$ and theirs duals $\delta _\alpha
=(\delta _i,\delta _a)$ (equivalently to an associated
N--connection structure), adapted to a symmetric metric field
$g_{\alpha \beta }$ (\ref{dm}) of necessary signature and to a
linear, in general nonsymmetric, connection $\Gamma _{~\beta
\gamma }^\alpha $ defining the covariant derivation $D_\alpha $
satisfying the metricity conditions $D_\alpha g_{\beta \gamma
}=0.$ The term la-- points to a prescribed type of anholonomy
structure.  As a matter of principle, on a (pseudo) Riemannian
spacetime, we can always, at least locally, remove our
considerations with respect to a coordinate basis. In this case
the geometric anisotopy is modelled by metrics of type
(\ref{ansatz}). Such ansatz for metrics are largely applied in
modern Kaluza--Klein theory \cite{over} where the N--conection
structures have been not pointed out because in the simplest
approximation on topological compactification of extra dimensions
the N--connection geometry is trivial. A rigorous anlysis of
systems with mixed holonomic--anholonomic variables was not yet
provided for general relativity,  extra dimension and gauge like
gravity theories..

A $n+m$ anholonomic structure distinguishes (d) the geometrical
objects into h-- and v--components. Such objects are briefly
called d--tensors, d--metrics and/or d--connections. Their
components are defined with respect to a la--basis of type
(\ref{dder}), its dual (\ref{ddif}), or their tensor products
(d--linear or d--affine transforms of such frames could also be
considered). For instance, a covariant and contravariant d--tensor
$Z,$ is expressed
\begin{equation}
\nonumber Z = Z_{~\beta }^\alpha \delta _\alpha \otimes \delta
^\beta = Z_{~j}^i\delta _i\otimes d^j+Z_{~a}^i\delta _i\otimes
\delta ^a+Z_{~j}^b\partial _b\otimes d^j+Z_{~a}^b\partial
_b\otimes \delta ^a.
\end{equation}

A linear d--connection $D$ on la--space $V^{(n+m)}{\cal ,}$ $$
D_{\delta _\gamma }\delta _\beta =\Gamma _{~\beta \gamma }^\alpha
\left( x,y\right) \delta _\alpha , $$ is parametrized by
non--trivial h--v--com\-po\-nents,
\begin{equation}
\label{dcon}\Gamma _{~\beta \gamma }^\alpha =\left(
L_{~jk}^i,L_{~bk}^a,C_{~jc}^i,C_{~bc}^a\right) .
\end{equation}

A metric on $V^{(n+m)}$ with $(m \times m)+ (n \times n)$
 block coefficients (\ref{dm}) is written in
distinguished form, as a metric d--tensor (in brief, d--metric),
with respect to a la--base (\ref{ddif})
\begin{equation}
\label{dmetric}\delta s^2 = g_{\alpha \beta }\left( u\right)
\delta ^\alpha \otimes \delta^\beta =
g_{ij}(x,y)dx^idx^j+h_{ab}(x,y)\delta y^a\delta y^b.
\end{equation}

Some d--connection and d--metric structures are compatible if
there are satisfied the conditions $$ D_\alpha g_{\beta \gamma
}=0. $$ For instance, a canonical compatible d--connection $$
^c\Gamma _{~\beta \gamma }^\alpha =\left(
^cL_{~jk}^i,^cL_{~bk}^a,^cC_{~jc}^i,^cC_{~bc}^a\right) $$ is
defined by the coefficients of d--metric (\ref{dmetric}),
$g_{ij}\left(
x,y\right) $ and $h_{ab}\left( x,y\right) ,$ and by the N--coefficients,%
\begin{eqnarray}
^cL_{~jk}^i & = & \frac 12g^{in}\left( \delta _kg_{nj}+\delta
_jg_{nk}-\delta _ng_{jk}\right) , \label{cdcon} \\ ^cL_{~bk}^a & =
& \partial _bN_k^a+\frac 12h^{ac}\left( \delta
_kh_{bc}-h_{dc}\partial _bN_i^d-h_{db}\partial _cN_i^d\right) ,
\nonumber \\ ^cC_{~jc}^i & = & \frac 12g^{ik}\partial _cg_{jk},
\nonumber \\ ^cC_{~bc}^a & = & \frac 12h^{ad}\left( \partial
_ch_{db}+\partial _bh_{dc}-\partial _dh_{bc}\right)  \nonumber
\end{eqnarray}
The coefficients of the canonical d--connection generalize for
la--spacetimes the well known Cristoffel symbols; on a (pseudo)
Riemannian spacetime with a fixed anholonomic frame the
d--connection coefficients transform exactly into the metric
connection coefficients.

For a d--connection (\ref{dcon}) the components of torsion,
\begin{eqnarray}
&T\left( \delta _\gamma ,\delta _\beta \right) &=T_{~\beta \gamma
}^\alpha \delta _\alpha ,  \nonumber \\ &T_{~\beta \gamma }^\alpha
&= \Gamma _{~\beta \gamma }^\alpha -\Gamma _{~\gamma \beta
}^\alpha +w_{~\beta \gamma }^\alpha \nonumber
\end{eqnarray}
are expressed via d--torsions
\begin{eqnarray}
T_{.jk}^i & = & - T_{.kj}^i=L_{jk}^i-L_{kj}^i,\quad
T_{ja}^i=C_{.ja}^i,T_{aj}^i=-C_{ja}^i, \nonumber \\ T_{.ab}^i & =
& 0,\quad T_{.bc}^a=S_{.bc}^a=C_{bc}^a-C_{cb}^a, \label{dtors} \\
T_{.ij}^a & = & -\Omega _{ij}^a,\quad T_{.bi}^a= \partial _b
N_i^a -L_{.bj}^a,\quad T_{.ib}^a=-T_{.bi}^a. \nonumber
\end{eqnarray}
We note that for symmetric linear connections the d--torsions are
induced as a pure anholonomic effect. They vanish with respect to
a coordinate frame of reference.

In a similar manner, putting non--vanishing coefficients
(\ref{dcon}) into the formula for curvature,
\begin{eqnarray}
&R\left( \delta _\tau ,\delta _\gamma \right) \delta _\beta &=
R_{\beta ~\gamma\tau }^{~\alpha }\delta _\alpha , \nonumber \\
&R_{\beta ~\gamma \tau }^{~\alpha } & =  \delta _\tau \Gamma
_{~\beta \gamma }^\alpha -\delta _\gamma \Gamma _{~\beta \delta
}^\alpha + \Gamma _{~\beta \gamma }^\varphi \Gamma _{~\varphi \tau
}^\alpha -\Gamma _{~\beta \tau }^\varphi \Gamma _{~\varphi \gamma
}^\alpha + \Gamma _{~\beta \varphi }^\alpha w_{~\gamma \tau
}^\varphi,  \nonumber
\end{eqnarray}
we can compute the components of d--curvatures
\begin{eqnarray}
R_{h.jk}^{.i} & = & \delta _kL_{.hj}^i-\delta_jL_{.hk}^i
 +  L_{.hj}^mL_{mk}^i-L_{.hk}^mL_{mj}^i-C_{.ha}^i\Omega _{.jk}^a,
\nonumber \\ R_{b.jk}^{.a} & = & \delta
_kL_{.bj}^a-\delta_jL_{.bk}^a
 +  L_{.bj}^cL_{.ck}^a-L_{.bk}^cL_{.cj}^a-C_{.bc}^a\Omega _{.jk}^c,
\nonumber \\ P_{j.ka}^{.i} & = & \partial _kL_{.jk}^i
+C_{.jb}^iT_{.ka}^b
  -  ( \partial _kC_{.ja}^i+L_{.lk}^iC_{.ja}^l -
L_{.jk}^lC_{.la}^i-L_{.ak}^cC_{.jc}^i ), \nonumber \\
P_{b.ka}^{.c} & = & \partial _aL_{.bk}^c +C_{.bd}^cT_{.ka}^d
  - ( \partial _kC_{.ba}^c+L_{.dk}^{c\,}C_{.ba}^d
- L_{.bk}^dC_{.da}^c-L_{.ak}^dC_{.bd}^c ),  \nonumber \\
S_{j.bc}^{.i} & = & \partial _cC_{.jb}^i-\partial _bC_{.jc}^i
 +  C_{.jb}^hC_{.hc}^i-C_{.jc}^hC_{hb}^i, \nonumber \\
S_{b.cd}^{.a} & = &\partial _dC_{.bc}^a-\partial
_cC_{.bd}^a+C_{.bc}^eC_{.ed}^a-C_{.bd}^eC_{.ec}^a. \nonumber
\end{eqnarray}

The Ricci tensor $$ R_{\beta \gamma }=R_{\beta ~\gamma \alpha
}^{~\alpha } $$ has the d--components
\begin{eqnarray}
R_{ij} & = & R_{i.jk}^{.k},\quad
 R_{ia}=-^2P_{ia}=-P_{i.ka}^{.k},\label{dricci} \\
R_{ai} &= & ^1P_{ai}=P_{a.ib}^{.b},\quad R_{ab}=S_{a.bc}^{.c}.
\nonumber
\end{eqnarray}
We point out that because, in general, $^1P_{ai}\neq ~^2P_{ia},$
the Ricci d-tensor is non symmetric.

Having defined a d-metric of type (\ref{dmetric}) in $V^{(n+m)}$
we can compute the scalar curvature $$ \overleftarrow{R}=g^{\beta
\gamma }R_{\beta \gamma } $$
of a d-connection $D,$%
\begin{equation}
\label{dscalar}{\overleftarrow{R}}=\widehat{R}+S,
\end{equation}
where $\widehat{R}=g^{ij}R_{ij}$ and $S=h^{ab}S_{ab}.$

Now, by introducing the values (\ref{dricci}) and (\ref{dscalar})
into the Einstein's equations $$ R_{\beta \gamma }-\frac
12g_{\beta \gamma }\overleftarrow{R}=k\Upsilon _{\beta \gamma },
$$ we can write down the system of field equations for la--gravity
with
 anholonomic (N--connection) structure:%
\begin{eqnarray}
R_{ij}-\frac 12\left( \widehat{R}+S\right) g_{ij} & = & k\Upsilon
_{ij}, \label{einsteq2} \\ S_{ab}-\frac 12\left(
\widehat{R}+S\right) h_{ab} & = & k\Upsilon _{ab},
 \nonumber \\
^1P_{ai} & = & k\Upsilon _{ai}, \nonumber \\ ^2P_{ia} & = &
-k\Upsilon _{ia}, \nonumber
\end{eqnarray}
where $\Upsilon _{ij},\Upsilon _{ab},\Upsilon _{ai}$ and $\Upsilon
_{ia}$ are the components of the energy--momentum d--tensor field
$\Upsilon _{\beta \gamma }$ (which includes possible cosmological
constants, contributions of anholonomy d--torsions (\ref{dtors})
and matter) and $k$ is the coupling constant.

The h- v- decomposition of gravitational field equations
(\ref{einsteq2}) was introduced by Miron and Anastasiei
\cite{ma94} in their N--connection approach to generalized Finsler
and Lagrange spaces. It holds true as well on (pseudo) Riemannian
spaces, in general gravity; in this case we obtain the usual form
of Einstein equations if we transfer considerations with respect
to coordinate frames. If the N--coefficients are prescribed by
fixing the anholonomic frame of reference, different classes of
solutions are to be constructed by finding the h-- and
v--components, $g_{ij}$ and $h_{ab},$ of metric (\ref{ansatz}), or
its equivalent (\ref{dm}). A more general approach is to consider
the N--connection as 'free' but subjected to the condition that
its coefficients along with the d--metric components are chosen to
solve the Einsten equations in the form (\ref{einsteq2}) for some
suggested symmetries, configurations of horizons and type of
singularities and well defined Cauchy problem. This way one can
construct new classes of metrics with generic local anisotropy
(see \cite{v1} and \cite{v2} and  Sections 7 in this paper).

\section{Higher Order Anisotropic Structures}

Miron and Atanasiu \cite{mat,m97,m97a} developed the higher order
Lagrange and Finsler geometry with applications in mechanics in
order to geometrize the concepts of classical mechanics on higher
order tangent bundles. The work \cite{vnp} was a proof that higher
order anisotropies (in brief, one writes abbreviations like ha--,
ha--superspace, ha--spacetime, ha--geometry and so on) can be
induced alternatively in low energy limits of (super) string
theories and a higher order superbundle N--connection formalism
was proposed. There were developed the theory of spinors \cite
{vjhep}, proposed models of ha--(super)gravity and matter
interactions on ha--spaces and defined the supersymmetric
stochastic calculus in ha--superspaces which were summarized in
the monograph \cite{vhp} containing a local (super) geometric
approach to so called ha--superstring and generalized
Finsler--Kaluza--Klein (super) gravities.

The aim of this section is to proof that higher order anisotropic
(ha--structu\-res) are induced by respective anholonomic frames in
higher dimension Einstein gravity, to present the basic geometric
background for a such moving frame formalism and associated
N--connections and to deduce the system of gravitational field
equations with respect to ha--frames.

\subsection{Ha--frames and corresponding N--connections}

Let us consider a (pseudo) Riemannian spacetime $V^{(\overline{n})}=V^{(n+%
\overline{m})}$ where the anisot\-rop\-ic dimension $\overline{m}$
is split
into $z$ sub--dimensions $m_p,$ $(p=1,2,...,z),$ i. e. $\overline{m}%
=m_1+m_2+...+m_z.$ The local coordinates on a such higher
dimension curved
spacetime will be denoted as to take into account the $m$--decomposition,%
\begin{eqnarray}
u&=&\{u^{\overline \alpha}\equiv u^{\alpha
_z}=(x^i,y^{a_1},y^{a_2},\ldots ,y^{a_p},\ldots y^{a_z})\},
\nonumber \\ u^{\alpha _p}&=&(x^i,y^{a_1},y^{a_2},\ldots
,y^{a_p})=\left( u^{\alpha _{p-1}},y^{a_p}\right) . \nonumber
\end{eqnarray}
The la--constructions from the previous Section are considered to
describe anholonomic structures of first order; for $z=1$ we put
$u^{\alpha _1}=\left( x^i,y^{a_1}\right) =u^\alpha =\left(
x^i,y^{a_1}\right) .$ The higher order anisotropies are defined
inductively, 'shell by shell', starting from the first order to
the higher order, $z$--anisotropy. In order
to distinguish the components of geometrical objects with respect to a $p$%
--shell we provide both Greek and Latin indices with a
corresponding subindex like ${\alpha _p}=({\alpha _{p-1}},{a_p}),$
and $a_p=(1,2,...,m_p),$ i. e. one holds a shell parametrization
for coordinates, $$
y^{a_p}=(y_{(p)}^1=y^1,y_{(p)}^2=y^2,...,y_{(p)}^{m_p}=y^{m_p}).
$$
We shall overline some indices, for instance, $\overline{\alpha }$ and $%
\overline{a},$ if would be necessary to point that it could be
split into shell components and omit the $p$--shell mark $(p)$ if
this does not lead to misunderstanding. Such decompositions of
indices and geometrical and physical values are introduced with
the aim for a further modelling of (in general, dynamical)
spllittings of higher dimension spacetimes, step by step, with
'interior' subspaces being of different dimension, to lower
dimensions, with nontrival topology and anholonomic (anisotropy)
structures in generalized Kaluza--Klein theories.

The coordinate frames are denoted $$
\partial _{\overline{\alpha }}=\partial /u^{\overline{\alpha }}=\left(
\partial /x^i,\partial /y^{a_1},...,\partial /y^{a_z}\right)
$$
with the dual ones%
$$ d^{\overline{\alpha }}=du^{\overline{\alpha }}=\left(
dx^i,dy^{a_1},...,dy^{a_z}\right) , $$ when $$
\partial _{\alpha _p}=\partial /u^{\alpha _p}=\left( \partial /x^i,\partial
/y^{a_1},...,\partial /y^{a_p}\right) $$ and $$ d^{\alpha
_p}=du^{\alpha _p}=\left( dx^i,dy^{a_1},...,dy^{a_p}\right) $$ if
considerations are limited to the $p$-th shell.

With respect to a coordinate frame a nonsigular metric%
$$
ds^2=\widetilde{g}_{\overline{\alpha }\overline{\beta }}\ du^{\overline{%
\alpha }}\otimes du^{\overline{\beta }} $$ with coefficients
$\widetilde{g}_{\overline{\alpha }\overline{\beta }}$ defined on
induction,
\begin{eqnarray}
\label{ansatz1}\widetilde{g}_{\alpha _1\beta _1}&=&\left[
\begin{array}{cc}
g_{ij}+M_i^{a_1}M_j^{b_1}h_{a_1b_1} & M_j^{e_1}h_{a_1e_1} \\
M_i^{e_1}h_{b_1e_1} & h_{a_1b_1}
\end{array}
\right] , \\
 & \vdots & \nonumber \\
\widetilde{g}_{\alpha _p\beta _p} &=&\left[
\begin{array}{cc}
g_{\alpha _{p-1}\beta _{p-1}}+M_{\alpha _{p-1}}^{a_p}M_{\beta
_{p-1}}^{b_p}h_{a_pb_p} & M_{\beta _{p-1}}^{e_p}h_{a_pe_p} \\
M_{\alpha _{p-1}}^{e_p}h_{b_pe_p} & h_{a_pb_p}
\end{array}
\right] , \nonumber \\
 &  \vdots &\nonumber \\
\widetilde{g}_{{\overline \alpha}{\overline \beta}} =
\widetilde{g}_{\alpha _z\beta _z} &=&\left[
\begin{array}{cc}
g_{\alpha _{z-1}\beta _{z-1}}+M_{\alpha _{z-1}}^{a_z}M_{\beta
_{z-1}}^{b_z}h_{a_zb_z} & M_{\beta _{z-1}}^{e_z}h_{a_ze_z} \\
M_{\alpha _{z-1}}^{e_z}h_{b_ze_z} & h_{a_zb_z}
\end{array}
\right] ,  \nonumber
\end{eqnarray}
where indices are split as $\alpha _1=\left( i_1,a_1\right) ,$
$\alpha _2=\left( \alpha _1,a_2\right) ,$ $\alpha _p=\left( \alpha
_{p-1},a_p\right) ;\ p=1,2,...z.$

The metric (\ref{ansatz1}) on $V^{(\overline{n})}$ splits into
symmetric blocks of matrices of dimensions $$ \left( n\times
n\right) \oplus \left( m_1\times m_1\right) \oplus ...\oplus
\left( m_z\times m_z\right) , $$ $n+m$ form
\begin{equation}
\label{dm1}g_{\alpha \beta }=\left(
\begin{array}{cccc}
g_{ij}(u) & 0 & \ldots & 0 \\ 0 & h_{a_1b_1} & \ldots & 0 \\
\ldots & \ldots & \cdots & \ldots \\ 0 & 0 & \ldots & h_{a_zb_z}
\end{array}
\right)
\end{equation}
with respect to an anholonomic frame basis defined on induction
\begin{eqnarray}\label{dder1}
\delta _{\alpha _p}&=&(\delta _{\alpha _{p-1}},\partial
_{a_p})=\left( \delta _i,\delta _{a_1},...,\delta
_{a_{p-1}},\partial _{a_p}\right)   \\
 &=&\frac \delta {\partial u^{\alpha _p}}=
 \left( \frac \delta {\partial
u^{\alpha _{p-1}}}=\frac \partial {\partial u^{\alpha
_{p-1}}}-N_{\alpha _{p-1}}^{b_p}\left( u\right) \frac \partial
{\partial y^{b_p}},\frac
\partial {\partial y^{a_p}}\right) , \nonumber
\end{eqnarray}
and
\begin{eqnarray} \label{ddif1}
\delta ^{\beta _p} &= &\left( d^i,\delta ^{{\overline a}_p}\right)
=\left( d^i,\delta ^{a_1},...,\delta ^{a_{p-1}},\delta
^{a_p}\right)  \\
 &=& \delta u^{\beta _p} = \left( d^i=dx^i,\delta ^{{\overline a}_p}=\delta
y^{{\overline a}_p}=dy^{{\overline a}_p}+ M_{\alpha
_{p-1}}^{{\overline a}_p}\left( u\right) du^{\alpha _{p-1}}\right)
, \nonumber
\end{eqnarray}
where $\overline{a}_p=\left( a_1,a_2,...,a_p\right) ,$ are called
the locally anisotropic bases (in brief la--bases) adapted
respectively to the N--coefficients $$ N_{\alpha
_{p-1}}^{a_p}=\left\{
N_i^{a_p},N_{a_1}^{a_p},...,N_{a_{p-2}}^{a_p},N_{a_{p-1}}^{a_p}\right\}
$$
and M--coefficients%
$$ M_{\alpha _{p-1}}^{a_p}=\left\{
M_i^{a_p},M_{a_1}^{a_p},...,M_{a_{p-2}}^{a_p},M_{a_{p-1}}^{a_p}\right\}
; $$ the coefficients $M_{\alpha _{p-1}}^{a_p}$ are related via
some algebraic relations with $N_{\alpha _{p-1}}^{a_p}$ in order
to be satisfied the la--basis duality conditions $$ \delta
_{\alpha _p}\otimes \delta ^{\beta _p}={\delta }_{\alpha
_p}^{\beta _p}, $$ where ${\delta }_{\alpha _p}^{\beta _p}$ is the
Kronecker symbol, for every shell.

The geometric structure of N-- and M--coefficients of a higher
order nonlinear connection becames more explicit if we write the
relations (\ref
{dder1}) and (\ref{ddif1}) in matrix form, respectively,%
$$ \delta _{\bullet }=\widehat{N}\left( u\right) \times \partial
_{\bullet } $$ and $$ \delta ^{\bullet }=d^{\bullet }\times
M\left( u\right) , $$
where%
$$ \delta _{\bullet }=\delta _{\overline \alpha}=\left(
\begin{array}{c}
\delta _i \\ \delta _{a_1} \\ \delta _{a_2} \\ \cdots \\ \delta
_{a_z}
\end{array}
\right) =\left(
\begin{array}{c}
\delta /\partial x^i \\ \delta /\partial y^{a_1} \\ \delta
/\partial y^{a_2} \\ \cdots \\ \delta /\partial y^{a_z}
\end{array}
\right) ,\partial _{\bullet }=\partial _{\overline \alpha}=\left(
\begin{array}{c}
\partial _i \\
\partial _{a_1} \\
\partial _{a_2} \\
\cdots \\
\partial _{a_z}
\end{array}
\right) =\left(
\begin{array}{c}
\partial /\partial x^i \\
\partial /\partial y^{a_1} \\
\partial /\partial y^{a_2} \\
\cdots \\
\partial /\partial y^{a_z}
\end{array}
\right) , $$ $$ \delta ^{\bullet }=\left(
\begin{array}{ccccc}
dx^i & \delta y^{a_1} & \delta y^{a_2} & \ldots & \delta y^{a_z}
\end{array}
\right) ,\ d^{\bullet }=\left(
\begin{array}{ccccc}
dx^i & dy^{a_1} & dy^{a_2} & \ldots & dy^{a_z}
\end{array}
\right) , $$ and $$ \widehat{N}=\left(
\begin{array}{ccccc}
1 & -N_i^{a_1} & -N_i^{a_2} & \ldots & -N_i^{a_z} \\ 0 & 1 &
-N_{a_1}^{a_2} & \ldots & -N_{a_1}^{a_z} \\ 0 & 0 & 1 & \ldots &
-N_{a_2}^{a_z} \\ \ldots & \ldots & \ldots & \ldots & \ldots \\ 0
& 0 & 0 & \ldots & 1
\end{array}
\right) , $$ $$ \ M=\left(
\begin{array}{ccccc}
1 & M_i^{a_1} & M_i^{a_2} & \ldots & M_i^{a_z} \\ 0 & 1 &
M_{a_1}^{a_2} & \ldots & M_{a_1}^{a_z} \\ 0 & 0 & 1 & \ldots &
M_{a_2}^{a_z} \\ \ldots & \ldots & \ldots & \ldots & \ldots \\ 0 &
0 & 0 & \ldots & 1
\end{array}
\right) . $$

The $n\times n$ matrice $g_{ij}$ defines the horizontal metric (in brief, $h$%
--metric) and the $m_p\times m_p$ matrices $h_{a_pb_p}$ defines
the vertical, $v_p$--metrics with respect to the associated
nonlinear connection (N--connection) structure given by its
coefficients $N_{\alpha _{p-1}}^{a_p}$ from (\ref{dder1}). The
geometry of N--connections on higher order tangent bundles is
studied in detail in \cite{mat,m97,m97a}, for vector
(super)bundles there it was proposed the approach from
\cite{vnp,vhp}; the approach and denotations elaborated in this
work is adapted to further applications in higher dimension
Einstein gravity and its non--Riemannian locally anisotropic
extensions.

A ha--basis $\delta _{\overline \alpha }$ (\ref{ddif}) on $V^{(\overline{n}%
)} $ is characterized by its anholonomy relations
\begin{equation}
\label{anholon1}\delta _{\overline{\alpha }}\delta _{\overline{\beta }%
}-\delta _{\overline{\beta }}\delta _{\overline{\alpha }}=w_{~\overline{%
\alpha }\overline{\beta }}^{\overline{\gamma }}\delta
_{\overline{\gamma }}.
\end{equation}
with anholonomy coefficients $w_{~\overline{\alpha }\overline{\beta }}^{%
\overline{\gamma }}.$ The anholonomy coefficients are computed
\begin{eqnarray}
w_{~ij}^k & = &
0;w_{~{a_p}j}^k=0;w_{~i{a_p}}^k=0;w_{~{a_p}{b_p}}^k=0;
w_{~{a_p}{b_p}}^{c_p}=0; \nonumber\\ w_{~ij}^{a_p} & = & -\Omega
_{ij}^{a_p}; w_{~{a_p}j}^{b_p}=-\delta _{a_p} N_i^{b_p};
 w_{~i{a_p}}^{b_p}=\delta _{a_p} N_i^{b_p};
\nonumber \\
w_{~{a_p}{b_p}}^{k_p} & = & 0;w_{~{a_p}{b_f}}^{c_f}=0, f<p%
; w_{~{b_f}{a_p}}^{c_f}=0, f<p; w_{~{a_p}{b_p}}^{c_f}=0, f<p;
\nonumber\\ w_{~{c_f}{d_s}}^{a_p} & = & -\Omega
_{{c_f}{d_s}}^{a_p}, (f,s<p); w_{~{a_p}{c_f}}^{b_p}=-\delta _{a_p}
N_{c_f}^{b_p}, f<p; w_{~{c_f}{a_p}}^{b_p}=\delta _{a_p}
N_{c_f}^{b_p}, f<p; \nonumber
\end{eqnarray}
where%
\begin{eqnarray}
\Omega _{ij}^{a_p}&=&\partial _iN_j^{a_p}-\partial
_jN_i^{a_p}+N_i^{b_p}\delta _{b_p}N_j^{a_p}-
 N_j^{b_p}\delta _{b_p}N_i^{a_p}, \label{ncurv1} \\
\Omega _{\alpha _f\beta _s}^{a_p} &=& \partial _{\alpha
_f}N_{\beta _s}^{a_p}-\partial _{\beta _s}N_{\alpha
_f}^{a_p}+N_{\alpha _f}^{b_p}\delta _{b_p}N_{\beta
_s}^{a_p}-N_{\beta_s}^{b_p} \delta _{b_p}N_{\alpha _f}^{a_p},
\nonumber
\end{eqnarray}
for $1\leq s,f<p,$ are the coefficients of higher order
N--connection curvature (N--curvature).

A higher order N--connection $N$ defines a global decomposition $$
N:\ V^{(\overline{n})}=H^{(n)}\oplus V^{(m_1)}\oplus
V^{(m_2)}\oplus ...\oplus V^{(m_z)}, $$ of spacetime
$V^{(\overline{n})}$ into a $n$--dimensional horizontal
subspace $H^{(n)}$ (with holonomic $x$--components) and into $m_p$%
--dimensional vertical subspaces $V^{(m_p)}$ (with anisotropic,
anholonomic, $y_{(p)}$--components).

\subsection{Distinguished linear connections}

In this section we consider fibered (pseudo) Riemannian manifolds $V^{(%
\overline{n})}$ enabled with anholonomic frame structures of basis
vector fields, $\delta ^{\overline{\alpha }}=(\delta ^i,\delta
^{\overline{a}})$
and theirs duals $\delta _{\overline{\alpha }}=(\delta _i,\delta _{\overline{%
a}})$ with associated N--connection structure, adapted to a
symmetric metric field $g_{\overline{\alpha }\overline{\beta }}$
(\ref{dm1}) and to a linear,
in general nonsymmetric, connection $\Gamma _{~\overline{\beta }\overline{%
\gamma }}^{\overline{\alpha }}$ defining the covariant derivation $D_{%
\overline{\alpha }}$ satisfying the metricity conditions $D_{\overline{%
\alpha }}g_{\overline{\beta }\overline{\gamma }}=0.$ Such
spacetimes are provided with anholonomic higher order anisotropic
structures and, in brief, are called ha--spacetimes.

A higher order N--connection distinguishes (d) the geometrical
objects into h-- and $v_p$--components (d--tensors, d--metrics
and/or d--connections). For instance, a d-tensor field of type
$\left(
\begin{array}{cccccc}
p & r_1 & ... & r_p & ... & r_z \\ q & s_1 & ... & s_p & ... & s_z
\end{array}
\right) $ is written in local form as%
\begin{eqnarray}
{\bf t} &=&t_{j_1...j_qb_1^{(1)}...b_{r_1}^{(1)}...b_1^{(p)}
...b_{r_p}^{(p)}...b_1^{(z)}...
b_{r_z}^{(z)}}^{i_1...i_pa_1^{(1)}...a_{r_1}^{(1)}...
a_1^{(p)}...a_{r_p}^{(p)}...a_1^{(z)}...a_{r_z}^{(z)}}\left(
u\right)
 \delta _{i_1}\otimes ...\otimes \delta _{i_p}\otimes d^{j_1}\otimes ...
\otimes d^{j_q}\otimes  \nonumber \\
 &{}&
\delta _{a_1^{(1)}}\otimes ...\otimes \delta
_{a_{r_1}^{(1)}}\otimes \delta ^{b_1^{(1)}}...\otimes \delta
^{b_{s_1}^{(1)}}\otimes ...\otimes \delta _{a_1^{(p)}}\otimes
...\otimes \delta _{a_{r_p}^{(p)}}\otimes ...\otimes
 \nonumber \\
 &{}&
\delta ^{b_1^{(p)}}...\otimes \delta ^{b_{s_p}^{(p)}}\otimes
\delta _{a_1^{(z)}}\otimes ...\otimes \delta
_{a_{rz}^{(z)}}\otimes \delta ^{b_1^{(z)}}...\otimes \delta
^{b_{s_z}^{(z)}}. \nonumber
\end{eqnarray}

A linear d--connection $D$ on ha--spacetime $V^{(\overline{n})},$
$$
D_{\delta _{\overline{\gamma }}}\delta _{\overline{\beta }}=\Gamma _{~%
\overline{\beta }\overline{\gamma }}^{\overline{\alpha }}\left(
u\right) \delta _{\overline{\alpha }}, $$ is defined by its
non--trivial h--v--components,
\begin{equation}
\label{dcon1}\Gamma _{~\overline{\beta }\overline{\gamma }}^{\overline{%
\alpha }}=\left( L_{~jk}^i,L_{~\overline{b}k}^{\overline{a}},C_{~j\overline{c%
}}^i,C_{~\overline{b}\overline{c}}^{\overline{a}%
},K_{~b_pc_p}^{a_p},K_{~b_sc_f}^{a_p},Q_{~b_fc_p}^{a_f}\right) ,
\end{equation}
for $f<p,s.$

A metric with block coefficients (\ref{dm1}) is written as a
d--metric, with respect to a la--base (\ref{ddif1})
\begin{equation}
\label{dmetric1}\delta s^2=g_{\overline{\alpha }\overline{\beta
}}\left(
u\right) \delta ^{\overline{\alpha }}\otimes \delta ^{\overline{\beta }%
}=g_{ij}(u)dx^idx^j+h_{a_pb_p}(u)\delta y^{a_p}\delta y^{b_p},
\end{equation}
where $p=1,2,...,z.$

A d--connection and a d--metric structure are compatible if there
are satisfied the conditions $$ D_{\overline{\alpha
}}g_{\overline{\beta }\overline{\gamma }}=0. $$

The canonical d--connection $^c\Gamma _{~\overline{\beta }\overline{\gamma }%
}^{\overline{\alpha }}$ is defined by the coefficients of
d--metric (\ref
{dmetric1}), and by the higher order N--coefficients,%
\begin{eqnarray}
^cL_{~jk}^i & = & \frac 12g^{in}\left( \delta _kg_{nj}+\delta
_jg_{nk}-\delta _ng_{jk}\right) , \label{cdcon1} \\
^cL_{~{\overline b}k}^{\overline a} & = & \delta _{\overline
b}N_k^{\overline a}+
 \frac 12h^{{\overline a}{\overline c}}
 \left( \partial _k h_{{\overline b}{\overline c}}-
h_{{\overline d}{\overline c}}
 \delta _{\overline b} N_k^{\overline d}
-h_{{\overline d}{\overline b}}\delta _{\overline c}N_k^{\overline
d}\right), \nonumber \\ ^cC_{~j{\overline c}}^i & = & \frac
12g^{ik}\delta _{\overline c}g_{jk}, \nonumber \\
^cC_{~{{\overline b}{\overline c}}}^{\overline a} & = & \frac
12h^{{\overline a}{\overline d}}\left( \delta _{\overline
c}h_{{\overline d}{\overline b}}+ \delta_{\overline
b}h_{{\overline d}{\overline c}}-\delta _{\overline d}
 h_{{\overline b}{\overline c}}\right),  \nonumber\\
^cK_{~{b_p}{c_p}}^{a_p} & = & \frac 12 g^{{a_p}{e_p}} \left(
\delta _{c_p} g_{{e_p}{b_p}}+ \delta _{b_p}g_{e_p c_p}-
\delta_{e_p} g_{b_p c_p}\right) , \nonumber \\
^cK_{~{b_s}e_f}^{a_p} & = & \delta _{b_s}N_{e_f}^{a_p}+
 \frac 12h^{a_p c_p}\left( \partial _{e_f} h_{b_s c_p}
-h_{d_p c_p}\delta _{b_s} N_{e_f}^{d_p} -h_{d_s b_s}\delta
_{c_p}N_{e_f}^{d_s}\right) , \nonumber \\ ^cQ_{~b_f c_p}^{a_f} & =
& \frac 12 h^{a_f e_f}\delta _{c_p}h_{b_f  e_f},
 \nonumber
\end{eqnarray}
where $f<p,s.$ They transform into usual Christoffel symbols with
respect to a coordinate base.

\subsection{Ha--torsions and ha--curvatures}

For a higher order anisotropic d--connection (\ref{dcon1}) the
components of torsion,
\begin{eqnarray}
&T\left( \delta _{\overline \gamma} ,\delta _{\overline \beta}
\right) &= T_{~{\overline \beta}{\overline \gamma} }^{\overline
\alpha} \delta _{\overline \alpha} ,  \nonumber \\ &T_{{\overline
\beta} {\overline \gamma} }^{\overline \alpha} &= \Gamma
_{~{\overline \beta}{\overline \gamma}}^{\overline \alpha} -
\Gamma _{~{\overline \gamma}{\overline  \beta} }^{\overline
\alpha} +
 w_{~{\overline \beta}{\overline  \gamma} }^{\overline \alpha} \nonumber
\end{eqnarray}
are expressed via d--torsions
\begin{eqnarray}
T_{.jk}^i&=&-T_{.kj}^i=L_{jk}^i-L_{kj}^i, \quad
T_{j\overline{a}}^i=-T_{\overline{a}j}^i=C_{.j\overline{a}}^i,
\nonumber \\ T_{.\overline{a}\overline{b}}^i&=& 0,
\quad T_{.\overline{b}\overline{c}}^{\overline{a}}= %
S_{.\overline{b}\overline{c}}^{\overline{a}}=%
C_{\overline{b}\overline{c}}^{\overline{a}}-%
C_{\overline{c}\overline{b}}^{\overline{a}}, %
\label{dtors1} \\
T_{.ij}^{\overline{a}}& =& %
-\Omega _{ij}^{\overline{a}},\quad %
T_{.\overline{b}i}^{\overline{a}}= %
-T_{.\overline{b}i}^{\overline{a}}= %
\delta _{\overline{b}}N_i^{\overline{a}}-L_{.\overline{b}j}^{\overline{a}}, %
 \nonumber \\ T_{.b_fc_f}^{a_f}& =&
-T_{.c_fb_f}^{a_f}=K_{.b_fc_f}^{a_f}-K_{.c_fb_f}^{a_f}, \nonumber
\\ T_{.a_pb_s}^{a_f}&=& 0,\quad
T_{.b_fa_p}^{a_f}=-T_{.a_pb_f}^{a_f}=Q_{.b_fa_p}^{a_f},
 \nonumber \\
T_{.a_fb_f}^{a_p}& = & -\Omega _{.a_fb_f}^{a_p},\quad
T_{.b_sa_f}^{a_p}=-T_{.a_fb_s}^{a_p}=\delta
_{b_s}N_{a_f}^{a_p}-K_{.b_sa_f}^{a_p}. \nonumber
\end{eqnarray}

We note that for symmetric linear connections the d--torsion is
induced as a pure anholonomic effect.

In a similar manner, putting non--vanishing coefficients
(\ref{dcon}) into the formula for curvature,
\begin{eqnarray}
&R\left( \delta _{\overline{\tau }},\delta _{\overline{\gamma
}}\right) \delta _{\overline{\beta }} & =R_{\overline{\beta
}~\overline{\gamma }\overline{\tau }}^{~\overline{\alpha }}\delta
_{\overline{\alpha }},\nonumber \\ & R_{\overline{\beta
}~\overline{\gamma } \overline{\tau }}^{~\overline{\alpha }} &
=\delta _{\overline{\tau }}
\Gamma _{~\overline{\beta }\overline{\gamma }}^{\overline{\alpha }}-%
\delta _{\overline{\gamma }}\Gamma _{~\overline{\beta }
\overline{\delta }}^{\overline{\alpha }}+\Gamma _{~\overline{\beta
}\overline{\gamma }}^{\overline{\varphi }}\Gamma
_{~\overline{\varphi }\overline{\tau }}^{\overline{\alpha
}}-\Gamma _{~\overline{\beta } \overline{\tau
}}^{\overline{\varphi }}\Gamma _{~\overline{\varphi }
\overline{\gamma }}^{\overline{\alpha }}+\Gamma _{~\overline{\beta
} \overline{\varphi }}^{\overline{\alpha }}w_{~\overline{\gamma }
\overline{\tau
}}^{\overline{\varphi }}, \nonumber %
\end{eqnarray}  we can compute the
components of d--curvatures   \begin{eqnarray}
R_{h.jk}^{.i}&=&\delta _kL_{.hj}^i-\delta _jL_{.hk}^i+
L_{.hj}^mL_{mk}^i-L_{.hk}^mL_{mj}^i-C_{.h\overline{a}}^i
\Omega_{.jk}^{\overline{a}}, \label{curvaturesha} \\
R_{\overline{b}.jk}^{.\overline{a}} &= &   \delta
_kL_{.\overline{b}j}^{\overline{a}}-\delta
_jL_{.\overline{b}k}^{\overline{a}}+L_{.\overline{b}j}^{\overline{c}}L_{.\overline{c}k}^{\overline{a}}-
L_{.\overline{b}k}^{\overline{c}}L_{.\overline{c}j}^{\overline{a}}-C_{.\overline{b}\overline{c}}^{\overline{a}}\Omega
_{.jk}^{\overline{c}}, \nonumber \\  P_{j.k\overline{a}}^{.i} & =
&   \partial
_kL_{.jk}^i+C_{.j\overline{b}}^iT_{.k\overline{a}}^{\overline{b}}-(\partial
_kC_{.j\overline{a}}^i+L_{.lk}^iC_{.j\overline{a}}^l-L_{.jk}^lC_{.l\overline{a}}^i-L_{.\overline{a}k}^{\overline{c}}C_{.j\overline{c}}^i),
\nonumber \\ P_{\overline{b}.k\overline{a}}^{.\overline{c}} & = &
\delta
_{\overline{a}}L_{.\overline{b}k}^{\overline{c}}+C_{.\overline{b}\overline{d}}^{\overline{c}}T_{.k\overline{a}}^{\overline{d}}-
(\partial
_kC_{.\overline{b}\overline{a}}^{\overline{c}}+L_{.\overline{d}k}^{\overline{c}\,}
C_{.\overline{b}\overline{a}}^{\overline{d}}-
L_{.\overline{b}k}^{\overline{d}}C_{.\overline{d}\overline{a}}^{\overline{c}}-L_{.\overline{a}k}^{\overline{d}}C_{.\overline{b}\overline{d}}^{\overline{c}}),
\nonumber \\ S_{j.\overline{b}\overline{c}}^{.i} &=&   \delta
_{\overline{c}}C_{.j\overline{b}}^i-\delta
_{\overline{b}}C_{.j\overline{c}}^i+C_{.j\overline{b}}^hC_{.h\overline{c}}^i-C_{.j\overline{c}}^hC_{h\overline{b}}^i,
\nonumber \\
S_{\overline{b}.\overline{c}\overline{d}}^{.\overline{a}}&= &
\delta
_{\overline{d}}C_{.\overline{b}\overline{c}}^{\overline{a}}-\delta
_{\overline{c}}C_{.\overline{b}\overline{d}}^{\overline{a}}+C_{.\overline{b}\overline{c}}^{\overline{e}}C_{.\overline{e}\overline{d}}^{\overline{a}}-
C_{.\overline{b}\overline{d}}^{\overline{e}}C_{.\overline{e}\overline{c}}^{\overline{a}},
\nonumber \\  W_{b_f.c_fe_f}^{.a_f} &= & \delta
_{e_f}K_{.b_fc_f}^{a_f}-\delta
_{c_f}K_{.b_fe_f}^{a_f}+K_{.b_fc_f}^{h_f}K_{h_fe_f}^{a_f}
\nonumber \\  &{}& - K_{.b_fe_f}^{h_f}K_{h_fc_f}^{a_f}
-Q_{.b_fa_p}^{a_f}\Omega _{.c_fe_f}^{a_p}, \nonumber \\
W_{b_s.c_fe_f}^{.a_p} & = & \delta _{e_f}K_{.b_sc_f}^{a_p}-\delta
_{c_f}K_{.b_se_f}^{a_p}+K_{.b_sc_f}^{c_p}K_{.c_pe_f}^{a_p}
\nonumber \\
 &{}& - K_{.b_se_f}^{c_p}L_{.c_pc_f}^{a_p}
-K_{.b_sc_p}^{a_p}\Omega _{.c_fe_f}^{c_p},
 \nonumber \\
Z_{b_f.c_fe_f}^{.a_f} & = &
\partial _{e_p}K_{.b_fc_f}^{a_f}+Q_{.b_fb_p}^{a_f}T_{.c_fe_p}^{b_p} \nonumber
\\
&{}& -(\partial
_{c_f}Q_{.b_fe_p}^{a_f}+K_{.h_fc_f}^{a_f}Q_{.b_fc_p}^{h_f}-
K_{.b_fc_f}^{h_f}Q_{.h_fe_p}^{a_f}-K_{.e_pc_f}^{c_p}C_{.b_fc_p}^{a_f}),
 \nonumber \\
Z_{b_r.c_fe_p}^{.c_s} &  = & \delta
_{e_p}K_{.b_rc_f}^{c_s}+K_{.b_rd_f}^{c_s}T_{.c_fe_p}^{d_f}
\nonumber \\
 &{}& -(\partial _{c_f}C_{.b_re_p}^{c_s}+K_{.d_fc_f}^{c_s\,}C_{.b_re_p}^{d_f}
-K_{.b_rc_f}^{d_t}C_{.d_te_p}^{c_s}-
K_{.e_pc_f}^{d_t}C_{.b_rd_t}^{c_s}), \nonumber \\
Y_{b_f.c_pe_p}^{.a_f}  & = & \delta _{e_p}Q_{.b_fc_p}^{a_f}-\delta
_{c_p}Q_{.b_fe_p}^{a_f}+Q_{.b_fc_p}^{d_f}Q_{.d_fe_p}^{a_f}-
Q_{.b_fe_p}^{d_f}Q_{d_fc_p}^{a_f}. \nonumber
\end{eqnarray}
where $f<p,s,r,t.$

\subsection{Einstein equations with respect to ha--frames}

The Ricci tensor $$
R_{\overline{\beta }\overline{\gamma }}=R_{\overline{\beta }~\overline{%
\gamma }\overline{\alpha }}^{~\overline{\alpha }} $$ has the
d--components
\begin{eqnarray}
R_{ij} & = & R_{i.jk}^{.k},\quad
 R_{i\overline a}=-^2P_{i \overline a}=-P_{i.k \overline a}^{.k},
\label{dricci1} \\ R_{{\overline a}i} &= & ^1P_{{\overline a}i}=
P_{{\overline a}.i{\overline b}}^{.\overline b}, \quad
R_{{\overline a}{\overline b}}=S_{{\overline a}.{\overline b}
 {\overline c}}^{.{\overline c}} \nonumber \\
R_{b_f c_f} & = & W_{b_f . c_f a_f}^{.a_f},\quad
 R_{e_p b_f}=-^2P_{b_f e_p}=- Z_{b_f . a_f e_p}^{.a_f}, \nonumber \\
R_{b_r c_f} &= & ^1P_{b_r c_f}= Z_{b_r . c_f e_s}^{.e_s}.
\nonumber
\end{eqnarray}
The Ricci d-tensor is non symmetric.

If a higher order d-metric of type (\ref{dmetric1}) is defined in $V^{(%
\overline{n})},$ we can compute the scalar curvature $$
\overline{R}=g^{\overline{\beta }\overline{\gamma }}R_{\overline{\beta }%
\overline{\gamma }}. $$
of a d-connection $D,$%
\begin{equation}
\label{dscalar1}{\overline{R}}=\widehat{R}+\overline{S},
\end{equation}
where $\widehat{R}=g^{ij}R_{ij}$ and $\overline{S}=h^{\overline{a}\overline{b%
}}S_{\overline{a}\overline{b}}.$

The h-v parametrization of the gravitational field equations in
ha--spacetimes is obtained by introducing the values (\ref{dricci1}) and (%
\ref{dscalar1}) into the Einstein's equations $$
R_{\overline{\beta }\overline{\gamma }}-\frac 12g_{\overline{\beta }%
\overline{\gamma }}\overline{R}=k\Upsilon _{\overline{\beta }\overline{%
\gamma }}, $$ and written
\begin{eqnarray}
R_{ij}-\frac 12\left( \widehat{R}+ \overline S\right) g_{ij} & = &
k\Upsilon _{ij}, \label{einsteq3} \\ S_{{\overline a} \overline
b}- \frac 12\left( \widehat{R}+\overline S\right)
 h_{{\overline a}{\overline b}} & = & k\Upsilon _{{\overline a}{\overline b}},
 \nonumber \\
^1P_{{\overline a}i}  =  k\Upsilon _{{\overline a}i},
  \quad ^1P_{a_p b_f} & = & k \Upsilon _{a_p b_f} \nonumber \\
^2P_{i \overline a}  =  -k\Upsilon _{i \overline a},\quad ^2P_{a_s
b_f} & = & - k \Upsilon _{a_f b_p}, \nonumber
\end{eqnarray}
where $\Upsilon _{ij},\Upsilon _{\overline{a}\overline{b}},\Upsilon _{%
\overline{a}i},\Upsilon _{i\overline{a}},\Upsilon
_{a_pb_f},\Upsilon _{a_fb_p}$ are the h-v--components of the
energy--mo\-men\-tum d--tensor field $\Upsilon _{\overline{\beta
}\overline{\gamma }}$ (which includes
possible cosmological constants, contributions of anholonomy d--torsions (%
\ref{dtors1}) and matter) and $k$ is the coupling constant.

We note that, in general, the ha--torsions are not vanishing.
Nevetheless, for a (pseudo)--Riemannian spacetime with induced
anholonomic anisotropies it is not necessary to consider an
additional to (\ref{einsteq3}) system of equations for torsion
becouse in this case the torsion structure is an
 anholonomic effect wich becames trivial with respect to holonomic
frames of reference.

If a ha--spacetime structure is associated to a generic nonzero
torsion, we could consider additionally, for instance, as in
\cite{vg}, a system of
algebraic d--field equations with a source $S_{~\overline{\beta }\overline{%
\gamma }}^{\overline{\alpha }}$ for a locally anisotropic spin
density of matter (if we construct a variant of higher order
anisotropic Einstein--Cartan theory): $$
T_{~\overline{\alpha }\overline{\beta }}^{\overline{\gamma }}+2\delta _{~[%
\overline{\alpha }}^{\overline{\gamma }}T_{~\overline{\beta }]\overline{%
\delta }}^{\overline{\delta }}=\kappa S_{~\overline{\alpha }\overline{\beta }%
.}^{\overline{\gamma }} $$ In a more general case we have to
introduce some new constraints and/or dynamical equations for
torsions and nonlinear connections which are induced from (super)
string theory and/ or higher order anisotropic supergravity
\cite{vap,vnp}. Two variants of gauge dynamical field equations
with both frame like and torsion variables will be considered in
the Section 5 and 6 of this paper.

\section{Gauge Fields on Ha--Spaces}

This section is devoted to gauge field theories on spacetimes
provided with higher order anisotropic anholonomic frame
structures.

\subsection{Bundles on ha--spaces}

Let us consider a principal bundle $\left( {\cal P},\pi ,Gr,V^{(\overline{n}%
)}\right) $ over a ha--spacetime $V^{(\overline{n})}$ (${\cal P}$ and $V^{(%
\overline{n})}$ are called respectively the base and total spaces)
with the
structural group $Gr$ and surjective map $\pi :{\cal P}\rightarrow V^{(%
\overline{n})}$ (on geometry of bundle spaces see, for instance,
\cite
{bishop,ma94,p}). At every point $u=(x,y_{(1)},...$ $,y_{(z)})$ $\in V^{(%
\overline{n})}$ there is a vicinity ${\cal U}\subset
V^{(\overline{n})},u\in
{\cal U,}$ with trivializing ${\cal P}$ diffeomorphisms $f$ and $\varphi :$%
\begin{eqnarray}
f_{{\cal U}}:\ \pi ^{-1}\left( {\cal U}\right) &\rightarrow& {\cal U\times }%
Gr,\qquad f\left( p\right) =\left( \pi \left( p\right) ,\varphi
\left( p\right) \right) , \nonumber \\ \varphi _{{\cal U}}:\ \pi
^{-1}\left( {\cal U}\right) &\rightarrow& Gr,\varphi (pq)=\varphi
\left( p\right) q  \nonumber
\end{eqnarray}
for every group element $q\in Gr$ and point $~p\in {\cal P}.$ We
remark that
in the general case for two open regions%
$$
{\cal U,V}\subset V^{(\overline{n})}{\cal ,U\cap V}\neq \emptyset ,f_{{\cal %
U|}_p}\neq f_{{\cal V|}_p},\mbox{ even }p\in {\cal U\cap V.} $$

Transition functions $g_{{\cal UV}}$ are defined $$ g_{{\cal
UV}}:{\cal U\cap V\rightarrow }Gr,g_{{\cal UV}}\left( u\right)
=\varphi _{{\cal U}}\left( p\right) \left( \varphi _{{\cal
V}}\left( p\right) ^{-1}\right) ,\pi \left( p\right) =u. $$

Hereafter we shall omit, for simplicity, the specification of
trivializing regions of maps and denote, for example, $f\equiv
f_{{\cal U}},\varphi \equiv \varphi _{{\cal U}},$ $s\equiv
s_{{\cal U}},$ if this will not give rise to ambiguities.

Let $\theta \,$ be the canonical left invariant 1-form on $Gr$
with values
in algebra Lie ${\cal G}$ of group $Gr$ uniquely defined from the relation $%
\theta \left( q\right) =q,$ for every $q\in {\cal G,}$ and
consider a 1-form $\omega $ on ${\cal U}\subset
V^{(\overline{n})}$ with values in ${\cal G.}$
Using $\theta $ and $\omega ,$ we can locally define the connection form $%
\Theta $ in ${\cal P}$ as a 1-form:
\begin{equation}
\label{bundcon}\Theta =\varphi ^{*}\theta +Ad\ \varphi ^{-1}\left(
\pi ^{*}\omega \right)
\end{equation}
where $\varphi ^{*}\theta $ and $\pi ^{*}\omega $ are,
respectively, 1--forms induced on $\pi ^{-1}\left( {\cal U}\right)
$ and ${\cal P}$ by maps $\varphi $ and $\pi $ and $\omega
=s^{*}\Theta .$ The adjoint action on a form $\lambda $ with
values in ${\cal G}$ is defined as $$ \left( Ad~\varphi
^{-1}\lambda \right) _p=\left( Ad~\varphi ^{-1}\left( p\right)
\right) \lambda _p $$ where $\lambda _p$ is the value of form
$\lambda $ at point $p\in {\cal P}.$

Introducing a basis $\{\Delta _{\widehat{a}}\}$ in ${\cal G}$ (index $%
\widehat{a}$ enumerates the generators making up this basis), we
write the 1-form $\omega $ on $V^{(\overline{n})}$ as
\begin{equation}
\label{form1}\omega =\Delta _{\widehat{a}}\omega
^{\widehat{a}}\left(
u\right) ,~\omega ^{\widehat{a}}\left( u\right) =\omega _{\overline{\mu }}^{%
\widehat{a}}\left( u\right) \delta u^{\overline{\mu }}
\end{equation}
where $\delta u^{\overline{\mu }}=\left( dx^i,\delta
y^{\overline{a}}\right)
$ and the Einstein summation rule on indices $\widehat{a}$ and $\overline{%
\mu }$ is used. Functions $\omega _{\overline{\mu
}}^{\widehat{a}}\left( u\right) $ from (\ref{form1}) are called
the components of Yang-Mills fields on ha-spacetime
$V^{(\overline{n})}.$ Gauge transforms of $\omega $ can be
interpreted as transition relations for $\omega _{{\cal U}}$ and $\omega _{%
{\cal V}},$ when $u\in {\cal U\cap V,}$%
\begin{equation}
\label{transf}\left( \omega _{{\cal U}}\right) _u=\left( g_{{\cal UV}%
}^{*}\theta \right) _u+Ad~g_{{\cal UV}}\left( u\right) ^{-1}\left( \omega _{%
{\cal V}}\right) _u.
\end{equation}

To relate $\omega _{\overline{\mu }}^{\widehat{a}}$ with a
covariant derivation we shall consider a vector bundle $\Xi $
associated to ${\cal P}.$
Let $\rho :Gr\rightarrow GL\left( {\cal R}^s\right) $ and $\rho ^{\prime }:%
{\cal G}\rightarrow End\left( E^s\right) $ be, respectively,
linear representations of group $Gr$ and Lie algebra ${\cal G}$
(where ${\cal R}$ is the real number field$).$ Map $\rho $ defines
a left action on $Gr$ and associated vector bundle $\Xi =P\times
{\cal R}^s/Gr,~\pi _E: E\rightarrow
V^{(\overline{n})}.$ Introducing the standard basis $\xi _{\underline{%
i}}=\{\xi _{\underline{1}},\xi _{\underline{2}},...,\xi
_{\underline{s}}\}$
in ${\cal R}^s,$ we can define the right action on $P\times $ ${\cal R}%
^s,\left( \left( p,\xi \right) q=\left( pq,\rho \left(
q^{-1}\right) \xi
\right) ,q\in Gr\right) ,$ the map induced from ${\cal P}$%
$$ p:{\cal R}^s\rightarrow \pi _E^{-1}\left( u\right) ,\quad
\left( p\left( \xi \right) =\left( p\xi \right) Gr,\xi \in {\cal
R}^s,\pi \left( p\right) =u\right) $$ and a basis of local
sections $e_{\underline{i}}:U\rightarrow \pi _E^{-1}\left(
U\right) ,~e_{\underline{i}}\left( u\right) =s\left( u\right)
\xi _{\underline{i}}.$ Every section $\varsigma :V^{(\overline{n}%
)}\rightarrow \Xi $ can be written locally as $\varsigma
=\varsigma ^ie_i,\varsigma ^i\in C^\infty \left( {\cal U}\right)
.$ To every vector
field $X$ on $V^{(\overline{n})}$ and Yang-Mills field $\omega ^{\widehat{a}%
} $ on ${\cal P}$ we associate operators of covariant derivations:
\begin{equation}
\label{operat}\nabla _X\zeta =e_{\underline{i}}\left[ X\zeta ^{\underline{i}%
}+B\left( X\right) _{\underline{j}}^{\underline{i}}\zeta ^{\underline{j}%
}\right] ,\ B\left( X\right) =\left( \rho ^{\prime }X\right) _{\widehat{a}%
}\omega ^{\widehat{a}}\left( X\right) .
\end{equation}
The transform (\ref{transf}) and operators (\ref{operat}) are
interrelated
by these transition transforms for values $e_{\underline{i}},\zeta ^{%
\underline{i}},$ and $B_{\overline{\mu }}:$
\begin{eqnarray}
e_{\underline{i}}^{{\cal V}}\left( u\right) & = & \left[ \rho
g_{{\cal UV}}\left(u\right) \right]
_{\underline{i}}^{\underline{j}}e_{\underline{i}}^{{\cal U}},
~\zeta _{{\cal U}}^{\underline{i}}\left( u\right) = \left[ \rho
g_{{\cal UV}}\left( u\right) \right]
_{\underline{i}}^{\underline{j}} \zeta _{{\cal
V}}^{\underline{i}}, \label{transf1} \\ B_{\overline{\mu }}^{{\cal
V}}\left( u\right) &=& \left[ \rho g_{{\cal UV}}\left( u\right)
\right] ^{-1} \delta _{\overline{\mu }}\left[ \rho g_{{\cal UV}}
\left( u\right) \right] +\left[ \rho g_{{\cal UV}}\left( u\right)
\right] ^{-1}B_{\overline{\mu }}^{{\cal U}}\left( u\right) \left[
\rho g_{{\cal UV}}\left( u\right) \right] , \nonumber
\end{eqnarray}
where $B_{\overline{\mu }}^{{\cal U}}\left( u\right) =B^{\overline{\mu }%
}\left( \delta /du^{\overline{\mu }}\right) \left( u\right) .$

Using (\ref{transf1}), we can verify that the operator $\nabla _X^{{\cal U}%
}, $ acting on sections of $\pi _\Xi :\Xi \rightarrow
V^{(\overline{n})}$
according to definition (\ref{operat}), satisfies the properties%
\begin{eqnarray}
\nabla _{f_1X+f_2Y}^{{\cal U}} &= & f_1\nabla _X^{{\cal
U}}+f_2\nabla _X^{{\cal U}},~\nabla _X^{{\cal U}}\left( f\zeta
\right) =f\nabla _X^{{\cal U}}\zeta +\left( Xf\right) \zeta ,
\nonumber \\ \nabla _X^{{\cal U}}\zeta &=& \nabla _X^{{\cal
V}}\zeta ,\quad u\in {\cal U\cap V,}f_1,f_2\in C^\infty \left(
{\cal U}\right) . \nonumber
\end{eqnarray}

So, we can conclude that the Yang--Mills connection in the vector bundle $%
\pi _\Xi :\Xi \rightarrow V^{(\overline{n})}$ is not a general
one, but is
induced from the principal bundle $\pi :{\cal P}\rightarrow V^{(\overline{n}%
)}$ with structural group $Gr.$

The curvature ${\cal K}$ of connection $\Theta $ from
(\ref{bundcon}) is defined as
\begin{equation}
\label{bundcurv}{\cal K}=D\Theta ,~D=\widehat{H}\circ d
\end{equation}
where $d$ is the operator of exterior derivation acting on ${\cal
G}$-valued forms as $$
d\left( \Delta _{\widehat{a}}\otimes \chi ^{\widehat{a}}\right) =\Delta _{%
\widehat{a}}\otimes d\chi ^{\widehat{a}} $$ and $\widehat{H}\,$ is
the horizontal projecting operator acting, for example, on the
1-form $\lambda $ as $\left( \widehat{H}\lambda \right) _P\left(
X_p\right) =\lambda _p\left( H_pX_p\right) ,$ where $H_p$ projects
on the horizontal subspace $$ {\cal H}_p\in P_p\left[ X_p\in {\cal
H}_p\mbox{ is equivalent to }\Theta _p\left( X_p\right) =0\right]
. $$ We can express (\ref{bundcurv}) locally as
\begin{equation}
\label{bundcurv1}{\cal K}=Ad~\varphi _{{\cal U}}^{-1}\left( \pi ^{*}{\cal K}%
_{{\cal U}}\right)
\end{equation}
where
\begin{equation}
\label{bundcurv2}{\cal K}_{{\cal U}}=d\omega _{{\cal U}}+\frac
12\left[ \omega _{{\cal U}},\omega _{{\cal U}}\right] .
\end{equation}
The exterior product of ${\cal G}$-valued form (\ref{bundcurv2})
is defined as $$
\left[ \Delta _{\widehat{a}}\otimes \lambda ^{\widehat{a}},\Delta _{\widehat{%
b}}\otimes \xi ^{\widehat{b}}\right] =\left[ \Delta _{\widehat{a}},\Delta _{%
\widehat{b}}\right] \otimes \lambda ^{\widehat{a}}\bigwedge \xi ^{\widehat{b}%
}, $$
where the antisymmetric tensorial product is denoted $\lambda ^{\widehat{a}%
}\bigwedge \xi ^{\widehat{b}}=\lambda ^{\widehat{a}}\xi ^{\widehat{b}}-\xi ^{%
\widehat{b}}\lambda ^{\widehat{a}}.$

Introducing structural coefficients
$f_{\widehat{b}\widehat{c}}^{\quad \widehat{a}}$ of ${\cal G}$
satisfying $$
\left[ \Delta _{\widehat{b}},\Delta _{\widehat{c}}\right] =f_{\widehat{b}%
\widehat{c}}^{\quad \widehat{a}}\Delta _{\widehat{a}} $$ we can
rewrite (\ref{bundcurv2}) in a form more convenient for local
considerations:
\begin{equation}
\label{bundcurv3}{\cal K}_{{\cal U}}=\Delta _{\widehat{a}}\otimes {\cal K}_{%
\overline{\mu }\overline{\nu }}^{\widehat{a}}\delta u^{\overline{\mu }%
}\bigwedge \delta u^{\overline{\nu }}
\end{equation}
where%
$$ {\cal K}_{\overline{\mu }\overline{\nu
}}^{\widehat{a}}=\frac{\delta \omega _{\overline{\nu
}}^{\widehat{a}}}{\partial u^{\overline{\mu }}}-\frac{\delta
\omega _{\overline{\mu }}^{\widehat{a}}}{\partial u^{\overline{\nu
}}}+\frac
12f_{\widehat{b}\widehat{c}}^{\quad \widehat{a}}\left( \omega _{\overline{%
\mu }}^{\widehat{b}}\omega _{\overline{\nu }}^{\widehat{c}}-\omega _{%
\overline{\nu }}^{\widehat{b}}\omega _{\overline{\mu
}}^{\widehat{c}}\right) .. $$

This subsection ends by considering the problem of reduction of
the local an\-i\-sot\-rop\-ic gauge symmetries and gauge fields to
isotropic ones. For local trivial considerations we can consider
that with respect to holonomic frames the higher order anisotropic
Yang-Mills fields reduce to usual ones on (pseudo) Riemannian
spaces.

\subsection{Yang-Mills equations on ha-spaces}

Interior gauge symmetries are associated to semisimple structural
groups. On the principal bundle $\left( {\cal P},\pi
,Gr,V^{({\overline{n}})}\right) $ with nondegenerate Killing form
for semisimple group $Gr$ we can define the generalized bundle
metric
\begin{equation}
\label{tmetric}h_p\left( X_p,Y_p\right) =G_{\pi \left( p\right)
}\left( d\pi _PX_P,d\pi _PY_P\right) +K\left( \Theta _P\left(
X_P\right) ,\Theta _P\left( X_P\right) \right) ,
\end{equation}
where $d\pi _P$ is the differential of map $\pi :{\cal P}\rightarrow V^{({%
\overline{n}})}{\cal ,}$ $G_{\pi \left( p\right) }$ is locally
generated as
the ha-metric (\ref{dmetric1}), and $K$ is the Killing form on ${\cal G:}$%
$$
K\left( \Delta _{\widehat{a}},\Delta _{\widehat{b}}\right) =f_{\widehat{b}%
\widehat{d}}^{\quad \widehat{c}}f_{\widehat{a}\widehat{c}}^{\quad \widehat{d}%
}=K_{\widehat{a}\widehat{b}}. $$

Using the metric $g_{\overline{\alpha }\overline{\beta }}$ on $V^{({%
\overline{n}})}$ (respectively, $h_P\left( X_P,Y_P\right) $ on
${\cal P)}$ we can introduce operators $*_G$ and $\widehat{\delta
}_G$ acting in the space of forms on $V^{({\overline{n}})}$ ($*_H$
and $\widehat{\delta }_H$ acting on forms on ${\cal P)}).$ Let
$e_{\overline{\mu }}$ be an orthonormalized frame on ${\cal
U\subset }V^{({\overline{n}})},$ locally adapted to the
N--connection structure, i. .e. being related via some local
distinguisherd linear transforms to a ha--frame (\ref{dder1}) and
$e^{\overline{\mu }}$ be the adjoint coframe. Locally%
$$
G=\sum\limits_{\overline{\mu }}\eta \left( \overline{\mu }\right) e^{%
\overline{\mu }}\otimes e^{\overline{\mu }}, $$
where $\eta _{\overline{\mu }\overline{\mu }}=\eta \left( \overline{\mu }%
\right) =\pm 1,$ $\overline{\mu }=1,2,...,\overline{n},$ and the
Hodge
operator $*_G$ can be defined as $*_G:\Lambda ^{\prime }\left( V^{({%
\overline{n}})}\right) \rightarrow \Lambda ^{\overline{n}}\left( V^{({%
\overline{n}})}\right) ,$ or, in explicit form, as
\begin{equation}
\label{hodge}*_G\left( e^{\overline{\mu }_1}\bigwedge ...\bigwedge e^{%
\overline{\mu }_r}\right) =\eta \left( \overline{\nu }_1\right)
...\eta \left( \overline{\nu }_{\overline{n}-r}\right) \times
\end{equation}
$$ sign\left(
\begin{array}{ccccccc}
1 & 2 & \ldots & r & r+1 & \ldots & \overline{n} \\ \overline{\mu
}_1 & \overline{\mu }_2 & \ldots & \overline{\mu }_r &
\overline{\nu }_1 & \ldots & \overline{\nu }_{\overline{n}-r}
\end{array}
\right) e^{\overline{\nu }_1}\bigwedge ...\bigwedge e^{\overline{\nu }_{%
\overline{n}-r}}. $$
Next, we define the operator%
$$ *_G^{-1}=\eta \left( 1\right) ...\eta \left(
\overline{n}\right) \left( -1\right) ^{r\left(
\overline{n}-r\right) }*_G $$ and introduce the scalar product on
forms $\beta _1,\beta _2,...\subset
\Lambda ^r\left( V^{({\overline{n}})}\right) $ with compact carrier:%
$$ \left( \beta _1,\beta _2\right) =\eta \left( 1\right) ...\eta
\left( n_E\right) \int \beta _1\bigwedge *_G\beta _2. $$ The
operator $\widehat{\delta }_G$ is defined as the adjoint to $d$
associated to the scalar product for forms, specified for
$r$-forms as
\begin{equation}
\label{adjoint}\widehat{\delta }_G=\left( -1\right)
^r*_G^{-1}\circ d\circ *_G.
\end{equation}

We remark that operators $*_H$ and $\delta _H$ acting in the total space of $%
{\cal P}$ can be defined similarly to (\ref{hodge}) and
(\ref{adjoint}), but by using metric (\ref{tmetric}). Both these
operators also act in the space
of ${\cal G}$-valued forms:%
$$
*\left( \Delta _{\widehat{a}}\otimes \varphi ^{\widehat{a}}\right) =\Delta _{%
\widehat{a}}\otimes (*\varphi ^{\widehat{a}}), $$ $$
\widehat{\delta }\left( \Delta _{\widehat{a}}\otimes \varphi ^{\widehat{a}%
}\right) =\Delta _{\widehat{a}}\otimes (\widehat{\delta }\varphi ^{\widehat{a%
}}). $$

The form $\lambda $ on ${\cal P}$ with values in ${\cal G}$ is
called horizontal if $\widehat{H}\lambda =\lambda $ and
equivariant if $R^{*}\left( q\right) \lambda =Ad~q^{-1}\varphi
,~\forall g\in Gr,R\left( q\right) $ being the right shift on
${\cal P}.$ We can verify that equivariant and
horizontal forms also satisfy the conditions%
$$ \lambda =Ad~\varphi _{{\cal U}}^{-1}\left( \pi ^{*}\lambda
\right) ,\qquad \lambda _{{\cal U}}=S_{{\cal U}}^{*}\lambda , $$
$$ \left( \lambda _{{\cal V}}\right) _{{\cal U}}=Ad~\left(
g_{{\cal UV}}\left( u\right) \right) ^{-1}\left( \lambda _{{\cal
U}}\right) _u. $$

Now, we can define the field equations for curvature
(\ref{bundcurv1}) and connection (\ref{bundcon}):
\begin{equation}
\label{ym1}\Delta {\cal K}=0,
\end{equation}
\begin{equation}
\label{ym2}\nabla {\cal K}=0,
\end{equation}
where $\Delta =\widehat{H}\circ \widehat{\delta }_H.$ Equations
(\ref{ym1}) are similar to the well-known Maxwell equations and
for non-Abelian gauge fields are called Yang-Mills equations. The
structural equations (\ref{ym2}) are called the Bianchi
identities.

The field equations (\ref{ym1}) do not have a physical meaning
because they are written in the total space of the bundle $\Xi $
and not on the base anisotropic spacetime $V^{({\overline{n}})}.$
But this difficulty may be
obviated by projecting the mentioned equations on the base. The 1-form $%
\Delta {\cal K}$ is horizontal by definition and its equivariance
follows from the right invariance of metric (\ref{tmetric}). So,
there is a unique form $(\Delta {\cal K})_{{\cal U}}$ satisfying
$$
\Delta {\cal K=}Ad~\varphi _{{\cal U}}^{-1}\pi ^{*}(\Delta {\cal K})_{{\cal U%
}}. $$
The projection of (\ref{ym1}) on the base can be written as $(\Delta {\cal K}%
)_{{\cal U}}=0.$ To calculate $(\Delta {\cal K})_{{\cal U}},$ we
use the
equality \cite{bishop,pd}%
$$
d\left( Ad~\varphi _{{\cal U}}^{-1}\lambda \right) =Ad~~\varphi _{{\cal U}%
}^{-1}~d\lambda -\left[ \varphi _{{\cal U}}^{*}\theta ,Ad~\varphi _{{\cal U}%
}^{-1}\lambda \right] $$
where $\lambda $ is a form on ${\cal P}$ with values in ${\cal G.}$ For $r$%
-forms we have $$ \widehat{\delta }\left( Ad~\varphi _{{\cal
U}}^{-1}\lambda \right) =Ad~\varphi _{{\cal
U}}^{-1}\widehat{\delta }\lambda -\left( -1\right)
^r*_H\{\left[ \varphi _{{\cal U}}^{*}\theta ,*_HAd~\varphi _{{\cal U}%
}^{-1}\lambda \right] $$ and, as a consequence,
\begin{equation}
\label{aux1}\widehat{\delta }{\cal K}=Ad~\varphi _{{\cal U}}^{-1}\{\widehat{%
\delta }_H\pi ^{*}{\cal K}_{{\cal U}}+*_H^{-1}[\pi ^{*}\omega _{{\cal U}%
},*_H\pi ^{*}{\cal K}_{{\cal U}}]\} -*_H^{-1}\left[ \Theta ,Ad~\varphi _{%
{\cal U}}^{-1}*_H\left( \pi ^{*}{\cal K}\right) \right] .
\end{equation}
By using straightforward calculations in the adapted dual basis on
$\pi ^{-1}\left( {\cal U}\right) $ we can verify the equalities
\begin{equation}
\label{aux2}\left[ \Theta ,Ad~\varphi _{{\cal U}}^{-1}~*_H\left( \pi ^{*}%
{\cal K}_{{\cal U}}\right) \right] =0,\widehat{H}\delta _H\left( \pi ^{*}%
{\cal K}_{{\cal U}}\right) =\pi ^{*}\left( \widehat{\delta }_G{\cal K}%
\right) ,
\end{equation}
$$
*_H^{-1}\left[ \pi ^{*}\omega _{{\cal U}},*_H\left( \pi ^{*}{\cal K}_{{\cal U%
}}\right) \right] =\pi ^{*}\{*_G^{-1}\left[ \omega _{{\cal U}},*_G{\cal K}_{%
{\cal U}}\right] \}. $$
>From (\ref{aux1}) and (\ref{aux2}) one follows that
\begin{equation}
\label{aux3}\left( \Delta {\cal K}\right) _{{\cal U}}=\widehat{\delta }_G%
{\cal K}_{{\cal U}}+*_G^{-1}\left[ \omega _{{\cal U}},*_G{\cal K}_{{\cal U}%
}\right] .
\end{equation}

Taking into account (\ref{aux3}) and (\ref{adjoint}), we prove
that projection on the base of equations (\ref{ym1}) and
(\ref{ym2}) can be expressed respectively as
\begin{equation}
\label{ym3}*_G^{-1}\circ d\circ *_G{\cal K}_{{\cal
U}}+*_G^{-1}\left[ \omega _{{\cal U}},*_G{\cal K}_{{\cal
U}}\right] =0.
\end{equation}
$$ d{\cal K}_{{\cal U}}+\left[ \omega _{{\cal U}},{\cal K}_{{\cal
U}}\right] =0. $$

Equations (\ref{ym3}) (see (\ref{aux3})) are gauge--invariant because%
$$ \left( \Delta {\cal K}\right) _{{\cal U}}=Ad~g_{{\cal
UV}}^{-1}\left( \Delta {\cal K}\right) _{{\cal V}}. $$

By using formulas (\ref{bundcurv3})-(\ref{adjoint}) we can rewrite (\ref{ym3}%
) in coordinate form
\begin{equation}
\label{ym4}D_{\overline{\nu }}\left( G^{\overline{\nu }\overline{\lambda }}%
{\cal K}_{~\overline{\lambda }\overline{\mu }}^{\widehat{a}}\right) +f_{%
\widehat{b}\widehat{c}}^{\quad \widehat{a}}g^{\overline{v}\overline{\lambda }%
}\omega _{\overline{\lambda }}^{~\widehat{b}}{\cal K}_{~\overline{\nu }%
\overline{\mu }}^{\widehat{c}}=0,
\end{equation}
where $D_{\overline{\nu }}$ is a compatible with metric covariant
derivation on ha-spacetime (\ref{ym4}).

We point out that for our bundles with semisimple structural
groups the Yang-Mills equations (\ref{ym1}) (and, as a
consequence, their horizontal projections (\ref{ym3}), or
(\ref{ym4})) can be obtained by variation of the action
\begin{equation}
\label{action}I=\int {\cal K}_{~\overline{\mu }\overline{\nu }}^{\widehat{a}}%
{\cal K}_{~\overline{\alpha }\overline{\beta }}^{\widehat{b}}G^{\overline{%
\mu }\overline{\alpha }}g^{\overline{\nu }\overline{\beta }}K_{\widehat{a}%
\widehat{b}}\left| g_{\overline{\alpha }\overline{\beta }}\right|
^{1/2}dx^1...dx^n\delta y_{(1)}^1...\delta y_{(1)}^{m_1}...\delta
y_{(z)}^1...\delta y_{(z)}^{m_z}.
\end{equation}
Equations for extremals of (\ref{action}) have the form $$
K_{\widehat{r}\widehat{b}}g^{\overline{\lambda }\overline{\alpha }}g^{%
\overline{\kappa }\overline{\beta }}D_{\overline{\alpha }}{\cal K}_{~%
\overline{\lambda }\overline{\beta }}^{\widehat{b}}-K_{\widehat{a}\widehat{b}%
}g^{\overline{\kappa }\overline{\alpha }}g^{\overline{\nu }\overline{\beta }%
}f_{\widehat{r}\widehat{l}}^{\quad \widehat{a}}\omega _{\overline{\nu }}^{%
\widehat{l}}{\cal K}_{~\overline{\alpha }\overline{\beta
}}^{\widehat{b}}=0, $$ which are equivalent to ''pure'' geometric
equations (\ref{ym4}) (or (\ref
{ym3})) due to nondegeneration of the Killing form $K_{\widehat{r}\widehat{b}%
}$ for semisimple groups.

To take into account gauge interactions with matter fields
(sections of vector bundle $\Xi $ on $V^{({\overline{n}})}$) we
have to introduce a source 1--form ${\cal J}$ in equations
(\ref{ym1}) and to write them
\begin{equation}
\label{yms}\Delta {\cal K}={\cal J}
\end{equation}

Explicit constructions of ${\cal J}$ require concrete definitions
of the bundle $\Xi ;$ for example, for spinor fields an invariant
formulation of the Dirac equations on ha--spaces is necessary. We
omit spinor considerations in this paper (see \cite{vjmp,vjhep}).

\section{Gauge Ha-gravity}

A considerable body of work on the formulation of gauge
gravitational models on isotropic spaces is based on application
of nonsemisimple groups, for example, of Poincare and affine
groups, as structural gauge groups (see critical analysis and
original results in \cite{deh,vg,mielke,hgmn,wal,tseyt,pbo}). The
main impediment to developing such models is caused by the
degeneration of Killing forms for nonsemisimple groups, which make
it impossible to construct consistent variational gauge field
theories (functional (\ref {action}) and extremal equations are
degenerate in these cases). There are at least two possibilities
to get around the mentioned difficulty.\ The first is to realize a
minimal extension of the nonsemisimple group to a semisimple one,
similar to the extension of the Poincare group to the de Sitter
group considered in \cite{p,pd,tseyt}. The second possibility is
to introduce into consideration the bundle of adapted affine
frames on la-space
$V^{({\overline{n}})},$ to use an auxiliary nondegenerate bilinear form $a_{%
\widehat{a}\widehat{b}}$ instead of the degenerate Killing form $K_{\widehat{%
a}\widehat{b}}$ and to consider a ''pure'' geometric method,
illustrated in the previous section, of definition of gauge field
equations. Projecting on the base $V^{({\overline{n}})},$ we shall
obtain gauge gravitational field equations on ha-space having a
form similar to Yang-Mills equations.

The goal of this section is to prove that a specific
parametrization of components of the Cartan connection in the
bundle of adapted affine frames on $V^{({\overline{n}})}$
establishes an equivalence between Yang-Mills equations
(\ref{yms}) and Einstein equations (\ref{einsteq3}) on ha--spaces.

\subsection{Bundles of linear ha--frames}

Let $\left( X_{\overline{\alpha }}\right) _u=\left( X_i,X_{\overline{a}%
}\right) _u=\left( X_i,X_{a_1},...,X_{a_z}\right) _u$ be a frame
locally adapted to the N--conection structure at a point $u\in
V^{({\overline{n}})}.$
We consider a local right distinguished action of matrices%
$$ A_{\overline{\alpha }^{\prime }}^{\quad \overline{\alpha
}}=\left(
\begin{array}{cccc}
A_{i^{\prime }}^{\quad i} & 0 & ... & 0 \\ 0 & B_{a_1^{\prime
}}^{\quad a_1} & ... & 0 \\ ... & ... & ... & ... \\ 0 & 0 & ... &
B_{a_z^{\prime }}^{\quad a_z}
\end{array}
\right) \subset GL_{\overline{n}}= $$ $$ GL\left( n,{\cal
R}\right) \oplus GL\left( m_1,{\cal R}\right) \oplus ...\oplus
GL\left( m_z,{\cal R}\right) . $$ Nondegenerate matrices
$A_{i^{\prime }}^{\quad i}$ and $B_{j^{\prime }}^{\quad j},$
respectively, transform linearly $X_{i|u}$ into $X_{i^{\prime
}|u}=A_{i^{\prime }}^{\quad i}X_{i|u}$ and $X_{a_p^{\prime }|u}$ into $%
X_{a_p^{\prime }|u}=B_{a_p^{\prime }}^{\quad a_p}X_{a_p|u},$ where $X_{%
\overline{\alpha }^{\prime }|u}=A_{\overline{\alpha }^{\prime
}}^{\quad \overline{\alpha }}X_{\overline{\alpha }}$ is also an
adapted frame at the
same point $u\in V^{({\overline{n}})}.$ We denote by $La\left( V^{({%
\overline{n}})}\right) $ the set of all adapted frames $X_{\overline{\alpha }%
}$ at all points of $V^{({\overline{n}})}$ and consider the surjective map $%
\pi $ from $La\left( V^{({\overline{n}})}\right) $ to
$V^{({\overline{n}})}$ transforming every adapted frame
$X_{\overline{\alpha }|u}$ and point $u$ into the point $u.$ Every
$X_{\overline{\alpha }^{\prime }|u}$ has a unique
representation as $X_{\overline{\alpha }^{\prime }}=A_{\overline{\alpha }%
^{\prime }}^{\quad \overline{\alpha }}X_{\overline{\alpha
}}^{\left( 0\right) },$ where $X_{\overline{\alpha }}^{\left(
0\right) }$ is a fixed distinguished basis in tangent space
$T\left( V^{({\overline{n}})}\right) .$
It is obvious that $\pi ^{-1}\left( {\cal U}\right) ,{\cal U}\subset V^{({%
\overline{n}})},$ is bijective to ${\cal U}\times
GL_{\overline{n}}\left( {\cal R}\right) .$ We can transform
$La\left( V^{({\overline{n}})}\right) $
in a differentiable manifold taking $\left( u^{\overline{\beta }},A_{%
\overline{\alpha }^{\prime }}^{\quad \overline{\alpha }}\right) $
as a local coordinate system on $\pi ^{-1}\left( {\cal U}\right)
.$ Now, it is easy to verify that $$
{\cal {L}}a(V^{({\overline{n}})})=(La(V^{({\overline{n}})}),V^{({\overline{n}%
})},GL_{\overline n}({\cal R})) $$ is a principal bundle. We call
${\cal {L}}a(V^{({\overline{n}})})$ the bundle of linear adapted
frames on $V^{({\overline{n}})}.$

The next step is to identify the components of, for simplicity,
compatible
d-connection $\Gamma _{\overline{\beta }\overline{\gamma }}^{\overline{%
\alpha }}$ on $V^{({\overline{n}})},$ with the connection in ${\cal {L}}%
a(V^{({\overline{n}})})$%
\begin{equation}
\label{conb}\Theta _{{\cal U}}^{\widehat{a}}=\omega
^{\widehat{a}}=\{\omega _{\quad \overline{\lambda
}}^{\widehat{\alpha }\widehat{\beta }}\doteq \Gamma
_{\overline{\beta }\overline{\gamma }}^{\overline{\alpha }}\}.
\end{equation}
Introducing (\ref{conb}) in (\ref{aux3}), we calculate the local
1-form
\begin{equation}
\label{aux4}\left( \Delta {\cal R}^{(\Gamma )}\right) _{{\cal U}}=\Delta _{%
\widehat{\alpha }\widehat{\alpha }_1}\otimes (g^{\overline{\nu }\overline{%
\lambda }}D_{\overline{\lambda }}{\cal R}_{\quad \overline{\nu }\overline{%
\mu }}^{\widehat{\alpha }\widehat{\gamma }}+f_{\quad \widehat{\beta }%
\widehat{\delta }\widehat{\gamma }\widehat{\varepsilon }}^{\widehat{\alpha }%
\widehat{\gamma }}g^{\overline{\nu }\overline{\lambda }}\omega
_{\quad
\lambda }^{\widehat{\beta }\widehat{\delta }}{\cal R}_{\quad \overline{\nu }%
\overline{\mu }}^{\widehat{\gamma }\widehat{\varepsilon }})\delta u^{%
\overline{\mu }},
\end{equation}
where $$ \Delta _{\widehat{\alpha }\widehat{\beta }}=\left(
\begin{array}{cccc}
\Delta _{\widehat{i}\widehat{j}} & 0 & ... & 0 \\ 0 & \Delta
_{\widehat{a}_1\widehat{b}_1} & ... & 0 \\ ... & ... & ... & ...
\\ 0 & 0 & ... & \Delta _{\widehat{a}_z\widehat{b}_z}
\end{array}
\right) $$
is the standard distinguished basis in the Lie algebra of matrices ${{\cal {G}}l}%
_{\overline{n}}\left( {\cal R}\right) $ with $(\Delta _{\widehat{i} \widehat{%
k}}) _{jl}$ $= \delta _{ij}\delta _{kl}$ and $\left( \Delta _{\widehat{a}_p%
\widehat{c}_p}\right) _{b_pd_p} = \delta _{a_pb_p}\delta _{c_pd_p}
$
defining  the stand\-ard bas\-es in ${\cal {G}}l\left( {\cal R}^{%
\overline{n}}\right) .$ We have denoted the curvature of
connection (\ref {conb}), considered in (\ref{aux4}), as $$
{\cal R}_{{\cal U}}^{(\Gamma )}=\Delta _{\widehat{\alpha }\widehat{\alpha }%
_1}\otimes {\cal R}_{\quad \overline{\nu }\overline{\mu }}^{\widehat{\alpha }%
\widehat{\alpha }_1}X^{\overline{\nu }}\bigwedge X^{\overline{\mu
}}, $$
where ${\cal R}_{\quad \overline{\nu }\overline{\mu }}^{\widehat{\alpha }%
\widehat{\alpha }_1}=R_{\overline{\alpha }_1\quad \overline{\nu }\overline{%
\mu }}^{\quad \overline{\alpha }}$ (see curvatures
(\ref{curvaturesha})).

\subsection{Bundles of affine ha--frames and Einstein equations}

Besides the bundles ${\cal {L}}a\left( V^{({\overline{n}})}\right)
$ on ha-spacetime $V^{({\overline{n}})},$ there is another bundle,
the bundle of adapted affine frames with structural group
$Af_{n_E}\left( {\cal R}\right)
=GL_{n_E}\left( V^{({\overline{n}})}\right) $ $\otimes {\cal R}^{\overline{n}%
},$ which can be naturally related to the gravity models on
(pseudo)
Riemannian spaces. Because as a linear space the Lie Algebra $af_{\overline{n%
}}\left( {\cal R}\right) $ is a direct sum of ${{\cal {G}}l}_{\overline{n}%
}\left( {\cal R}\right) $ and ${\cal R}^{\overline{n}},$ we can
write forms on ${\cal {A}}a\left( V^{({\overline{n}})}\right) $ as
$\Phi =\left( \Phi
_1,\Phi _2\right) ,$ where $\Phi _1$ is the ${{\cal {G}}l}_{\overline{n}%
}\left( {\cal R}\right) $ component and $\Phi _2$ is the ${\cal R}^{%
\overline{n}}$ component of the form $\Phi .$ The connection (\ref{conb}), $%
\Theta $ in ${{\cal {L}}a}\left( V^{({\overline{n}})}\right) ,$
induces the
Cartan connection $\overline{\Theta }$ in ${{\cal {A}}a}\left( V^{({%
\overline{n}})}\right) ;$ see the isotropic case in
\cite{p,pd,bishop}.
There is only one connection on ${{\cal {A}}a}\left( V^{({\overline{n}}%
)}\right) $ represented as $i^{*}\overline{\Theta }=\left( \Theta
,\chi
\right) ,$ where $\chi $ is the shifting form and $i:{{\cal {A}}a}%
\rightarrow {{\cal {L}}a}$ is the trivial reduction of bundles. If $s_{{\cal %
U}}^{(a)}$ is a local adapted frame in ${{\cal {L}}a}\left( V^{({\overline{n}%
})}\right) ,$ then $\overline{s}_{{\cal U}}^{\left( 0\right) }=i\circ s_{%
{\cal U}}$ is a local section in ${{\cal {A}}a}\left( V^{({\overline{n}}%
)}\right) $ and
\begin{equation}
\label{curvbaf}\left( \overline{\Theta }_{{\cal U}}\right) =s_{{\cal U}%
}\Theta =\left( \Theta _{{\cal U}},\chi _{{\cal U}}\right) ,
\end{equation}
where $\chi =e_{\widehat{\alpha }}\otimes \chi _{~\overline{\mu }}^{\widehat{%
\alpha }}X^{\overline{\mu }},$ $g_{\overline{\alpha
}\overline{\beta }}=\chi
_{~\overline{\alpha }}^{\widehat{\alpha }}\chi _{~\overline{\beta }}^{%
\widehat{\beta }}\eta _{\widehat{\alpha }\widehat{\beta }}\quad (\eta _{%
\widehat{\alpha }\widehat{\beta }}$ is diagonal with $\eta
_{\widehat{\alpha }\widehat{\alpha }}=\pm 1)$ is a frame
decomposition of metric (\ref {dmetric1}) on
$V^{({\overline{n}})},e_{\widehat{\alpha }}$ is the standard
distinguished basis on ${\cal R}^{\overline{n}},$ and the
projection of torsion , $T_{{\cal U}},$ on the base
$V^{({\overline{n}})}$ is defined as
\begin{equation}
\label{torsbaf}T_{{\cal U}}=d\chi _{{\cal U}}+\Omega _{{\cal
U}}\bigwedge
\chi _{{\cal U}}+\chi _{{\cal U}}\bigwedge \Omega _{{\cal U}}=e_{\widehat{%
\alpha }}\otimes \sum\limits_{\overline{\mu }\overline{\nu }}T_{~\overline{%
\mu }\overline{\nu }}^{\widehat{\alpha }}X^{\overline{\mu }}\bigwedge X^{%
\overline{\nu }}.
\end{equation}
For a fixed locally adapted basis on ${\cal U}\subset
V^{({\overline{n}})}$
we can identify components $T_{~\overline{\mu }\overline{\nu }}^{\widehat{a}%
} $ of torsion (\ref{torsbaf}) with components of torsion (\ref{dtors1}) on $%
V^{({\overline{n}})},$ i.e. $T_{~\overline{\mu }\overline{\nu }}^{\widehat{%
\alpha }}=T_{~\overline{\mu }\overline{\nu }}^{\overline{\alpha
}}.$ By straightforward calculation we obtain
\begin{equation}
\label{aux5}{(\Delta \overline{{\cal R}})}_{{\cal U}}=[{(\Delta {\cal R}%
^{(\Gamma )})}_{{\cal U}},\ {(R\tau )}_{{\cal U}}+{(Ri)}_{{\cal
U}}],
\end{equation}
where%
$$
\left( R\tau \right) _{{\cal U}}=\widehat{\delta }_GT_{{\cal U}%
}+*_G^{-1}\left[ \Omega _{{\cal U}},*_GT_{{\cal U}}\right] ,\quad
\left(
Ri\right) _{{\cal U}}=*_G^{-1}\left[ \chi _{{\cal U}},*_G{\cal R}_{{\cal U}%
}^{(\Gamma )}\right] . $$ Form $\left( Ri\right) _{{\cal U}}$ from
(\ref{aux5}) is locally constructed by using components of the
Ricci tensor (see (\ref{dricci1})) as follows
from decomposition on the local adapted basis $X^{\overline{\mu }}=\delta u^{%
\overline{\mu }}:$ $$
\left( Ri\right) _{{\cal U}}=e_{\widehat{\alpha }}\otimes \left( -1\right) ^{%
\overline{n}+1}R_{\overline{\lambda }\overline{\nu }}g^{\widehat{\alpha }%
\overline{\lambda }}\delta u^{\overline{\mu }}. $$

We remark that for isotropic torsionless pseudo-Riemannian spaces
the requirement that $\left( \Delta \overline{{\cal R}}\right)
_{{\cal U}}=0,$
i.e., imposing the connection (\ref{conb}) to satisfy Yang-Mills equations (%
\ref{ym1}) (equivalently (\ref{ym3}) or (\ref{ym4})) we obtain
\cite{p,pd} the equivalence of the mentioned gauge gravitational
equations with the vacuum Einstein equations $R_{ij}=0.\,$ In the
case of ha--spaces with arbitrary given torsion, even considering
vacuum gravitational fields, we have to introduce a source for
gauge gravitational equations in order to compensate for the
contribution of torsion and to obtain equivalence with the
Einstein equations.

Considerations presented in this section constitute the proof of
the following result:

\begin{theorem}
The Einstein equations (\ref{einsteq3}) for ha--gravity are
equivalent to the Yang-Mills equations
\begin{equation}
\label{yme}\left( \Delta \overline{{\cal R}}\right)
=\overline{{\cal J}}
\end{equation}
for the induced Cartan connection $\overline{\Theta }$ (see
(\ref{conb}) and
(\ref{curvbaf})) in the bundle of locally adapted affine frames ${\cal A}%
a\left( V^{({\overline{n}})}\right) $ with the source $\overline{{\cal J}}_{%
{\cal U}}$ constructed locally by using the same formulas (\ref{aux5}) for $%
\left( \Delta \overline{{\cal R}}\right) $, but where $R_{\overline{\alpha }%
\overline{\beta }}$ is changed by the matter source ${E}_{\overline{\alpha }%
\overline{\beta }}-\frac 12g_{\overline{\alpha }\overline{\beta }}{E}$ with $%
{E}_{\overline{\alpha }\overline{\beta }}=k\Upsilon _{\overline{\alpha }%
\overline{\beta }}-\lambda g_{\overline{\alpha }\overline{\beta
}}.$
\end{theorem}

We note that this theorem is an extension for higher order
anisotropic spacetimes of the Popov and Dikhin result \cite{pd}
with respect to a possible gauge like treatment of the Einstein
gravity. Similar theorems have been proved for locally anisotropic
gauge gravity \cite{vg} and in the framework of some variants of
locally (and higher order) anisotropic supergravity \cite {vhp}.

\section{Nonlinear De Sitter Gauge Ha--Gravity}

The equivalent reexpression of the Einstein theory as a gauge like
theory implies, for both locally isotropic and anisotropic
space--times, the nonsemisimplicity of the gauge group, which
leads to a nonvariational theory in the total space of the bundle
of locally adapted affine frames. A variational gauge
gravitational theory can be formulated by using a minimal
extension of the affine structural group ${{\cal
A}f}_{\overline{n}}\left( {\cal R}\right) $ to the de Sitter gauge
group $S_{\overline{n}}=SO\left( \overline{n}\right) $ acting on
distinguished ${\cal R}^{\overline{n}+1}$ space.

\subsection{Nonlinear gauge theories of de Sitter group}

Let us consider the de Sitter space $\Sigma ^{\overline{n}}$ as a
hypersurface given by the equations $\eta _{AB}u^Au^B=-l^2$ in the flat $%
\left( \overline{n}+1\right) $--dimensional space enabled with
diagonal
metric $\eta _{AB},\eta _{AA}=\pm 1$ (in this subsection $A,B,C,...=1,2,...,%
\overline{n}+1),(\overline{n}=n+m_1+...+m_z),$ where $\{u^A\}$ are
global Cartesian coordinates in ${\cal R}^{\overline{n}+1};l>0$ is
the curvature of de Sitter space. The de Sitter group $S_{\left(
\eta \right) }=SO_{\left( \eta \right) }\left(
\overline{n}+1\right) $ is defined as the isometry group of
$\Sigma ^{\overline{n}}$--space with $\frac{\overline{n}}2\left(
\overline{n}+1\right) $ generators of Lie algebra ${{\it
s}o}_{\left( \eta \right) }\left( \overline{n}+1\right) $
satisfying the commutation relations
\begin{equation}
\label{gener}\left[ M_{AB},M_{CD}\right] =\eta _{AC}M_{BD}-\eta
_{BC}M_{AD}-\eta _{AD}M_{BC}+\eta _{BD}M_{AC}.
\end{equation}

Decomposing indices $A,B,...$ as $A=\left( \widehat{\alpha },\overline{n}%
+1\right) ,B=\left( \widehat{\beta },\overline{n}+1\right) ,$
$...,$ the
metric $\eta _{AB}$ as $\eta _{AB}=\left( \eta _{\widehat{\alpha }\widehat{%
\beta }},\eta _{\left( \overline{n}+1\right) \left(
\overline{n}+1\right)
}\right) ,$ and operators $M_{AB}$ as $M_{\widehat{\alpha }\widehat{\beta }}=%
{\cal F}_{\widehat{\alpha }\widehat{\beta }}$ and $P_{\widehat{\alpha }%
}=l^{-1}M_{\overline{n}+1,\widehat{\alpha }},$ we can write
(\ref{gener}) as $$
\left[ {\cal F}_{\widehat{\alpha }\widehat{\beta }},{\cal F}_{\widehat{%
\gamma }\widehat{\delta }}\right] =\eta _{\widehat{\alpha }\widehat{\gamma }}%
{\cal F}_{\widehat{\beta }\widehat{\delta }}-\eta _{\widehat{\beta }\widehat{%
\gamma }}{\cal F}_{\widehat{\alpha }\widehat{\delta }}+\eta
_{\widehat{\beta
}\widehat{\delta }}{\cal F}_{\widehat{\alpha }\widehat{\gamma }}-\eta _{%
\widehat{\alpha }\widehat{\delta }}{\cal F}_{\widehat{\beta
}\widehat{\gamma }}, $$ $$
\left[ P_{\widehat{\alpha }},P_{\widehat{\beta }}\right] =-l^{-2}{\cal F}_{%
\widehat{\alpha }\widehat{\beta }},\quad \left[ P_{\widehat{\alpha }},{\cal F%
}_{\widehat{\beta }\widehat{\gamma }}\right] =\eta _{\widehat{\alpha }%
\widehat{\beta }}P_{\widehat{\gamma }}-\eta _{\widehat{\alpha }\widehat{%
\gamma }}P_{\widehat{\beta }}, $$ where we have indicated the
possibility to decompose ${{\it s}o}_{\left(
\eta \right) }\left( \overline{n}+1\right) $ into a direct sum, ${{\it s}o}%
_{\left( \eta \right) }\left( \overline{n}+1\right) ={{\it
s}o}_{\left( \eta \right) }(\overline{n})\oplus v_{\overline{n}},$
where $v_{\overline{n}}$ is the vector space stretched on vectors
$P_{\widehat{\alpha }}.$ We remark that $\Sigma
^{\overline{n}}=S_{\left( \eta \right) }/L_{\left( \eta \right)
},$ where $L_{\left( \eta \right) }=SO_{\left( \eta \right)
}\left(
\overline{n}\right) .$ For $\eta _{AB}=diag\left( 1,-1,-1,-1\right) $ and $%
S_{10}=SO\left( 1,4\right) ,L_6=SO\left( 1,3\right) $ is the group
of Lorentz rotations.

Let $W\left( {\cal E},{\cal R}^{\overline{n}+1},S_{\left( \eta \right) },%
{\cal P}\right) $ be the vector bundle associated with the principal bundle $%
{\cal P}\left( S_{\left( \eta \right) },{\cal E}\right) $ on ha-spacetime $%
v_{\overline{n}},$ where $S_{\left( \eta \right) }$ is taken to be
the structural group and by ${\cal E}$ it is denoted the total
space. The action of the structural group $S_{\left( \eta \right)
}$ on ${\cal E}\,$ can be realized by using $\overline{n}\times
\overline{n}$ matrices with a
parametrization distinguishing subgroup $L_{\left( \eta \right) }:$%
\begin{equation}
\label{aux6}B=bB_L,
\end{equation}
where%
$$ B_L=\left(
\begin{array}{cc}
L & 0 \\ 0 & 1
\end{array}
\right) , $$ $L\in L_{\left( \eta \right) }$ is the de Sitter bust
matrix transforming the vector $\left( 0,0,...,\rho \right) \in
{\cal R}^{\overline{n}+1}$ into
the point $\left( v^1,v^2,...,v^{\overline{n}+1}\right) \in \Sigma _\rho ^{%
\overline{n}}\subset {\cal R}^{\overline{n}+1}$ for which
$v_Av^A=-\rho ^2,v^A=t^A\rho .$ Matrix $b$ can be expressed $$
b=\left(
\begin{array}{cc}
\delta _{\quad \widehat{\beta }}^{\widehat{\alpha }}+\frac{t^{\widehat{%
\alpha }}t_{\widehat{\beta }}}{\left( 1+t^{\overline{n}+1}\right)
} & t^{ \widehat{\alpha }} \\ t_{\widehat{\beta }} &
t^{\overline{n}+1}
\end{array}
\right) . $$

The de Sitter gauge field is associated with a linear connection
in $W$,
i.e., with a ${{\it s}o}_{\left( \eta \right) }\left( \overline{n}+1\right) $%
-valued connection 1--form on $V^{({\overline{n}})}:$%
\begin{equation}
\label{convsit}\breve \Theta=\left(
\begin{array}{cc}
\omega _{\quad \widehat{\beta }}^{\widehat{\alpha }} & \breve
\theta^{ \widehat{\alpha }} \\ \breve \theta_{\widehat{\beta }} &
0
\end{array}
\right) ,
\end{equation}
where $\omega _{\quad \widehat{\beta }}^{\widehat{\alpha }}\in so(\overline{n%
})_{\left( \eta \right) },$ $\breve \theta^{\widehat{\alpha }}\in {\cal R}^{%
\overline{n}},\breve \theta_{\widehat{\beta }}\in \eta _{\widehat{\beta }%
\widehat{\alpha }}\breve \theta^{\widehat{\alpha }}.$

Because $S_{\left( \eta \right) }$-transforms mix $\omega _{\quad \widehat{%
\beta }}^{\widehat{\alpha }}$ and $\breve \theta ^{\widehat{\alpha
}}$ fields in (\ref{convsit}) (the introduced
para\-met\-ri\-za\-ti\-on is invariant on action on $SO_{\left(
\eta \right) }\left( \overline{n}\right) $
group we cannot identify $\omega _{\quad \widehat{\beta }}^{\widehat{\alpha }%
}$ and $\breve \theta ^{\widehat{\alpha }},$ respectively, with
the connection $\Gamma _{~\overline{\beta }\overline{\gamma
}}^{\overline{\alpha
}}$ and the fundamental form $\chi ^{\overline{\alpha }}$ in $V^{({\overline{%
n}})}$ (as we have for (\ref{conb}) and (\ref{curvbaf})). To avoid
this difficulty we consider \cite{tseyt,pbo} a nonlinear gauge
realization of the de Sitter group $S_{\left( \eta \right) }$ by
introducing the nonlinear gauge field
\begin{equation}
\label{convsitn}\Theta =b^{-1}\breve \Theta b+b^{-1}db=\left(
\begin{array}{cc}
\Gamma _{~\widehat{\beta }}^{\widehat{\alpha }} & \theta ^{
\widehat{\alpha }} \\ \theta _{\widehat{\beta }} & 0
\end{array}
\right) ,
\end{equation}
where $$
\Gamma _{\quad \widehat{\beta }}^{\widehat{\alpha }}=\omega _{\quad \widehat{%
\beta }}^{\widehat{\alpha }}-\left( t^{\widehat{\alpha }}Dt_{\widehat{\beta }%
}-t_{\widehat{\beta }}Dt^{\widehat{\alpha }}\right) /\left( 1+t^{\overline{n}%
+1}\right) , $$ $$
\theta ^{\widehat{\alpha }}=t^{\overline{n}+1}\breve \theta ^{\widehat{%
\alpha }}+Dt^{\widehat{\alpha }}-t^{\widehat{\alpha }}\left( dt^{\overline{n}%
+1}+\breve \theta _{\widehat{\gamma }}t^{\widehat{\gamma }}\right)
/\left( 1+t^{\overline{n}+1}\right) , $$ $$
Dt^{\widehat{\alpha }}=dt^{\widehat{\alpha }}+\omega _{\quad \widehat{\beta }%
}^{\widehat{\alpha }}t^{\widehat{\beta }}. $$

The action of the group $S\left( \eta \right) $ is nonlinear,
yielding transforms $$ \Gamma ^{\prime }=L^{\prime }\Gamma \left(
L^{\prime }\right) ^{-1}+L^{\prime }d\left( L^{\prime }\right)
^{-1},\theta ^{\prime }=L\theta , $$
where the nonlinear matrix-valued function $L^{\prime }=L^{\prime }\left( t^{%
\overline{\alpha }},b,B_T\right) $ is defined from $B_b=b^{\prime
}B_{L^{\prime }}$ (see the  parametrization (\ref{aux6})).

Now, we can identify components of (\ref{convsitn}) with components of $%
\Gamma _{~\overline{\beta }\overline{\gamma }}^{\overline{\alpha }}$ and $%
\chi _{~\overline{\alpha }}^{\widehat{\alpha }}$ on
$V^{({\overline{n}})}$
and induce in a consistent manner on the base of bundle $W( {\cal E},%
{\cal R}^{\overline{n}+1},S_{( \eta )},{\cal P})$ the
ha--geometry.

\subsection{Dynamics of the nonlinear de Sitter ha--gravity}

Instead of the gravitational potential (\ref{conb}), we introduce
the gravitational connection (similar to (\ref{convsitn}))
\begin{equation}
\label{convsitn1}\Gamma =\left(
\begin{array}{cc}
\Gamma _{~\widehat{\beta }}^{\widehat{\alpha }} & l_0^{-1}\chi ^{
\widehat{\alpha }} \\ l_0^{-1}\chi _{\widehat{\beta }} & 0
\end{array}
\right)
\end{equation}
where $$
\Gamma _{~\widehat{\beta }}^{\widehat{\alpha }}=\Gamma _{~\widehat{\beta }%
\overline{\mu }}^{\widehat{\alpha }}\delta u^{\overline{\mu }}, $$
$$ \Gamma _{\quad \widehat{\beta }\overline{\mu
}}^{\widehat{\alpha }}=\chi
_{\quad \overline{\alpha }}^{\widehat{\alpha }}\chi _{\quad \overline{\beta }%
}^{\widehat{\beta }}\Gamma _{\quad \overline{\beta }\overline{\gamma }}^{%
\overline{\alpha }}+\chi _{\quad \overline{\alpha }}^{\widehat{\alpha }%
}\delta _{\overline{\mu }}\chi _{\quad \widehat{\beta }}^{\overline{\alpha }%
}, $$
$\chi ^{\widehat{\alpha }}=\chi _{\quad \overline{\mu }}^{\widehat{\alpha }%
}\delta u^{\overline{\mu }},$ and $g_{\overline{\alpha }\overline{\beta }%
}=\chi _{\quad \overline{\alpha }}^{\widehat{\alpha }}\chi _{\quad \overline{%
\beta }}^{\widehat{\beta }}\eta _{\widehat{\alpha }\widehat{\beta }},$ and $%
\eta _{\widehat{\alpha }\widehat{\beta }}$ is parametrized as $$
\eta _{\widehat{\alpha }\widehat{\beta }}=\left(
\begin{array}{cccc}
\eta _{ij} & 0 & ... & 0 \\ 0 & \eta _{a_1b_1} & ... & 0 \\ ... &
... & ... & ... \\ 0 & 0 & ... & \eta _{a_zb_z}
\end{array}
\right) , $$ $\eta _{ij}=\left( 1,-1,...,-1\right) ,...\eta
_{ij}=\left( \pm 1,\pm 1,...,\pm 1\right) ,...,l_0$ is a
dimensional constant.

The curvature of (\ref{convsitn1}), ${\cal R}^{(\Gamma )}=d\Gamma
+\Gamma \bigwedge \Gamma ,$ can be written as
\begin{equation}
\label{curvdsit}{\cal R}^{(\Gamma )}=\left(
\begin{array}{cc}
{\cal R}_{\quad \widehat{\beta }}^{\widehat{\alpha }}+l_0^{-1}\pi _{\widehat{%
\beta }}^{\widehat{\alpha }} & l_0^{-1}T^{ \widehat{\alpha }} \\
l_0^{-1}T^{\widehat{\beta }} & 0
\end{array}
\right) ,
\end{equation}
where $$
\pi _{\widehat{\beta }}^{\widehat{\alpha }}=\chi ^{\widehat{\alpha }%
}\bigwedge \chi _{\widehat{\beta }},{\cal R}_{\quad \widehat{\beta }}^{%
\widehat{\alpha }}=\frac 12{\cal R}_{\quad \widehat{\beta }\overline{\mu }%
\overline{\nu }}^{\widehat{\alpha }}\delta u^{\overline{\mu
}}\bigwedge \delta u^{\overline{\nu }}, $$ and $$
{\cal R}_{\quad \widehat{\beta }\overline{\mu }\overline{\nu }}^{\widehat{%
\alpha }}=\chi _{\widehat{\beta }}^{\quad \overline{\beta }}\chi _{\overline{%
\alpha }}^{\quad \widehat{\alpha }}R_{\overline{\beta }.\overline{\mu }%
\overline{\nu }}^{~\overline{\alpha }} $$ (see
(\ref{curvaturesha}) for components of d-curvatures). The de
Sitter gauge group is semisimple and we are able to construct a
variational gauge
gravitational locally an\-iso\-trop\-ic theory (bundle metric (\ref{tmetric}%
) is nondegenerate). The Lagrangian of the theory is postulated as
$$ L=L_{\left( G\right) }+L_{\left( m\right) } $$ where the gauge
gravitational Lagrangian is defined as $$
L_{\left( G\right) }=\frac 1{4\pi }Tr\left( {\cal R}^{(\Gamma )}\bigwedge *_G%
{\cal R}^{(\Gamma )}\right) ={\cal L}_{\left( G\right) }\left|
g\right| ^{1/2}\delta ^{\overline{n}}u, $$
\begin{equation}
\label{lagrsit}{\cal L}_{\left( G\right) }=\frac 1{2l^2}T_{\quad \overline{%
\mu }\overline{\nu }}^{\widehat{\alpha }}T_{\widehat{\alpha
}}^{\quad
\overline{\mu }\overline{\nu }}+\frac 1{8\lambda }{\cal R}_{\quad \widehat{%
\beta }\overline{\mu }\overline{\nu }}^{\widehat{\alpha }}{\cal
R}_{\quad
\widehat{\alpha }}^{\widehat{\beta }\quad \overline{\mu }\overline{\nu }%
}-\frac 1{l^2}\left( {\overleftarrow{R}}\left( \Gamma \right)
-2\lambda _1\right) ,
\end{equation}
$T_{\quad \overline{\mu }\overline{\nu }}^{\widehat{\alpha }}=\chi
_{\quad \overline{\alpha }}^{\widehat{\alpha }}T_{\quad
\overline{\mu }\overline{\nu
}}^{\overline{\alpha }}$ (the gravitational constant $l^2$ in (\ref{lagrsit}%
) satisfies the relations $l^2=2l_0^2\lambda ,\lambda _1=-3/l_0],$
$\quad Tr$ denotes the trace on $\widehat{\alpha },\widehat{\beta
}$ indices, and the
matter field Lagrangian is defined as%
$$
L_{\left( m\right) }=\frac 12Tr\left( \Gamma \bigwedge *_G{\cal I}\right) =%
{\cal L}_{\left( m\right) }\left| g\right| ^{1/2}\delta
^{\overline{n}}u, $$
\begin{equation}
\label{lagrsitm}{\cal L}_{\left( m\right) }=\frac 12\Gamma _{\quad \widehat{%
\beta }\overline{\mu }}^{\widehat{\alpha }}S_{\quad \overline{\alpha }}^{%
\widehat{\beta }\quad \overline{\mu }}-t_{\quad \widehat{\alpha }}^{%
\overline{\mu }}l_{\quad \overline{\mu }}^{\widehat{\alpha }}.
\end{equation}
The matter field source ${\cal I}$ is obtained as a variational
derivation of ${\cal L}_{\left( m\right) }$ on $\Gamma $ and is
parametrized as
\begin{equation}
\label{sourcesit}{\cal I}=\left(
\begin{array}{cc}
S_{\quad \widehat{\beta }}^{\widehat{\alpha }} & -l_0t^{
\widehat{\alpha }} \\ -l_0t_{\widehat{\beta }} & 0
\end{array}
\right)
\end{equation}
with $t^{\widehat{\alpha }}=t_{\quad \overline{\mu }}^{\widehat{\alpha }%
}\delta u^{\overline{\mu }}$ and $S_{\quad \widehat{\beta }}^{\widehat{%
\alpha }}=S_{\quad \widehat{\beta }\overline{\mu }}^{\widehat{\alpha }%
}\delta u^{\overline{\mu }}$ being respectively the canonical
tensors of energy-momentum and spin density. Because of the
contraction of the ''interior'' indices $\widehat{\alpha
},\widehat{\beta }$ in (\ref{lagrsit}) and (\ref{lagrsitm}) we
used the Hodge operator $*_G$ instead of $*_H$ (hereafter we
consider $*_G=*).$

Varying the action $$ S=\int \left| g\right| ^{1/2}\delta
^{\overline{n}}u\left( {\cal L}_{\left( G\right) }+{\cal
L}_{\left( m\right) }\right) $$ on the $\Gamma $-variables
(\ref{convsitn1}), we obtain the gauge--gravitational field
equations:
\begin{equation}
\label{dsitteq1}d\left( *{\cal R}^{(\Gamma )}\right) +\Gamma
\bigwedge \left( *{\cal R}^{(\Gamma )}\right) -\left( *{\cal
R}^{(\Gamma )}\right) \bigwedge \Gamma =-\lambda \left( *{\cal
I}\right) .
\end{equation}

Specifying the variations on $\Gamma _{\quad \widehat{\beta }}^{\widehat{%
\alpha }}$ and $l^{\widehat{\alpha }}$-variables, we rewrite (\ref{dsitteq1}%
) as
\begin{equation}
\label{dsitteq2}\widehat{{\cal D}}\left( *{\cal R}^{(\Gamma )}\right) +\frac{%
2\lambda }{l^2}\left( \widehat{{\cal D}}\left( *\pi \right) +\chi
\bigwedge \left( *T^T\right) -\left( *T\right) \bigwedge \chi
^T\right) =-\lambda \left( *S\right) ,
\end{equation}
\begin{equation}
\label{dsitteq3}\widehat{{\cal D}}\left( *T\right) -\left( *{\cal R}%
^{(\Gamma )}\right) \bigwedge \chi -\frac{2\lambda }{l^2}\left(
*\pi \right) \bigwedge \chi =\frac{l^2}2\left( *t+\frac 1\lambda
*\tau \right) ,
\end{equation}
where $$
T^t=\{T_{\widehat{\alpha }}=\eta _{\widehat{\alpha }\widehat{\beta }}T^{%
\widehat{\beta }},~T^{\widehat{\beta }}=\frac 12T_{\quad \overline{\mu }%
\overline{\nu }}^{\widehat{\beta }}\delta u^{\overline{\mu
}}\bigwedge \delta u^{\overline{\nu }}\}, $$ $$
\chi ^T=\{\chi _{\widehat{\alpha }}=\eta _{\widehat{\alpha }\widehat{\beta }%
}\chi ^{\widehat{\beta }},~\chi ^{\widehat{\beta }}=\chi _{\quad \overline{%
\mu }}^{\widehat{\beta }}\delta u^{\overline{\mu }}\},\qquad \widehat{{\cal D%
}}=d+\widehat{\Gamma } $$
($\widehat{\Gamma }$ acts as $\Gamma _{\quad \widehat{\beta }\overline{\mu }%
}^{\widehat{\alpha }}$ on indices $\widehat{\gamma
},\widehat{\delta },...$ and as $\Gamma _{\quad \overline{\beta
}\overline{\mu }}^{\overline{\alpha }} $ on indices
$\overline{\gamma },$ $\overline{\delta },...).$ In (\ref
{dsitteq3}), $\tau $ defines the energy--momentum tensor of the
$S_{\left(
\eta \right) }$--gauge gravitational field $\widehat{\Gamma }:$%
\begin{equation}
\label{sourcesit1}\tau _{\overline{\mu }\overline{\nu }}\left( \widehat{%
\Gamma }\right) =\frac 12Tr\left( {\cal R}_{\overline{\mu }\overline{\alpha }%
}{\cal R}_{\quad \overline{\nu }}^{\overline{\alpha }}-\frac 14{\cal R}_{%
\overline{\alpha }\overline{\beta }}{\cal R}^{\overline{\alpha }\overline{%
\beta }} g_{\overline{\mu }\overline{\nu }}\right) .
\end{equation}

Equations (\ref{dsitteq1}) (or, equivalently, (\ref{dsitteq2}) and
(\ref {dsitteq3})) make up the complete system of variational
field equations for nonlinear de Sitter gauge gravity with higher
order anisotropy. They can be interpreted as a variant of gauge
like equations for ha--gravity \cite{vg} when the (pseudo)
Riemannian base frames and torsions are considered to be induced
by an anholonomic frame structure with associated N--connection

A. Tseytlin \cite{tseyt} presented a quantum analysis of the
isotropic version of equations (\ref{dsitteq2}) and
(\ref{dsitteq3}). Of course, the problem of quantizing
gravitational interactions is unsolved for both variants of
locally anisotropic and isotropic gauge de Sitter gravitational
theories, but we think that the generalized Lagrange version of
$S_{\left( \eta \right) }$-gravity is more adequate for studying
quantum radiational and statistical gravitational processes. This
is a matter for further investigations.

Finally, we remark that we can obtain a nonvariational Poincare
gauge gravitational theory on ha--spaces if we consider the
contraction of the gauge potential (\ref{convsitn1}) to a
potential with values in the Poincare Lie algebra $$ \Gamma
=\left(
\begin{array}{cc}
\Gamma _{\quad \widehat{\beta }}^{\widehat{\alpha }} &
l_0^{-1}\chi ^{ \widehat{\alpha }} \\ l_0^{-1}\chi
_{\widehat{\beta }} & 0
\end{array}
\right) \rightarrow \Gamma =\left(
\begin{array}{cc}
\Gamma _{\quad \widehat{\beta }}^{\widehat{\alpha }} &
l_0^{-1}\chi ^{ \widehat{\alpha }} \\ 0 & 0
\end{array}
\right) . $$ Isotropic Poincare gauge gravitational theories are
studied in a number of papers (see, for example,
\cite{wal,tseyt,pbo}). In a manner similar to considerations
presented in this work, we can generalize Poincare gauge models
for spaces with local anisotropy.

\section{An Ansatz for 4D d--Metrics}

We consider a 4D spacetime $V^{(3+1)}$ provided with a d--metric
(\ref {dmetric}) when $g_i = g_i (x^k)$ and $h_a = h_a (x^k, z)$
for $y^a = (z, y^4).$
 The N--connection coefficients are  some functions on three coordinates $(x^i,z),$%
\begin{eqnarray}
N_1^3&=&q_1(x^i,z),\ N_2^3=q_2(x^i,z), \label{ncoef} \\
N_1^4&=&n_1(x^i,z),\ N_2^4=n_2(x^i,z). \nonumber
\end{eqnarray}
For simplicity, we shall use brief denotations of partial derivatives, like $%
\dot a$$=\partial a/\partial x^1,a^{\prime }=\partial a/\partial x^2,$ $%
a^{*}=\partial a/\partial z$ $\dot a^{\prime }$$=\partial
^2a/\partial x^1\partial x^2,$ $a^{**}=\partial ^2a/\partial
z\partial z.$

The non--trivial components of the Ricci d--tensor (\ref{dricci}),
 for the mentioned type of d--metrics depending on three variables, are%
\begin{eqnarray}
&R_1^1&=R_2^2=\frac 1{2g_1g_2} [-(g_1^{^{\prime \prime }}+{\ddot
g}_2)+ \frac 1{2g_2}\left( {\dot g}_2^2+g_1^{\prime }g_2^{\prime
}\right) +
\frac 1{2g_1}\left( g_1^{\prime \ 2}+%
\dot g_1\dot g_2\right) ]; \label{ricci1} \\
\label{ricci2} &S_3^3&=S_4^4=\frac 1{h_3h_4}[-h_4^{**}+\frac 1{2h_4}(h_4^{*})^2+%
\frac 1{2h_3}h_3^{*}h_4^{*}]; \\ &P_{31}&=\frac{q_1}2[\left(
\frac{h_3^{*}}{h_3}\right) ^2-
\frac{h_3^{**}}{h_3}+%
\frac{h_4^{*}}{2h_4^{\ 2}}-\frac{h_3^{*}h_4^{*}}{2h_3h_4}]
+\frac 1{2h_4}[\frac{\dot h_4}{2h_4}h_4^{*}-\dot h_4^{*}+ %
\frac{\dot h_3}{2h_3}h_4^{*}],  \label{ricci3} \\ &{}& \nonumber
\\
&P_{32}&=\frac{q_2}2[\left( \frac{h_3^{*}}{h_3}\right) ^2-\frac{h_3^{**}}{h_3}+%
\frac{h_4^{*}}{2h_4^{\ 2}}-\frac{h_3^{*}h_4^{*}}{2h_3h_4}]
+\frac 1{2h_4}[\frac{h_4^{\prime }}{2h_4}h_4^{*}-h_4^{\prime \ *}+ %
\frac{h_3^{\prime }}{2h_3}h_4^{*}];  \nonumber \\ &{}&   \nonumber
\\ & P_{41}&=-\frac{h_4}{2h_3}n_1^{**} +
\frac{1}{4h_3}(\frac{h_4}{h_3} h^*_3 - 3 h^*_4) n^*_1 ,
\label{ricci4} \\ & P_{42}&= -\frac{h_4}{2h_3}n_2^{**} +
\frac{1}{4h_3}(\frac{h_4}{h_3} h^*_3 - 3 h^*_4) n^*_2. \nonumber
\end{eqnarray}

The curvature scalar $\overleftarrow{R}$ (\ref{dscalar}) is
defined by the sum of two non-trivial components
$\widehat{R}=2R_1^1$ and $S=2S_3^3.$

The system of Einstein equations (\ref{einsteq2}) transforms into
\begin{eqnarray}
R_1^1&=&-\kappa \Upsilon _3^3=-\kappa \Upsilon _4^4,
\label{einsteq3a} \\ S_3^3&=&-\kappa \Upsilon _1^1=-\kappa
\Upsilon _2^2, \label{einsteq3b}\\ P_{3i}&=& \kappa \Upsilon
_{3i}, \label{einsteq3c} \\ P_{4i}&=& \kappa \Upsilon _{4i},
\label{einsteq3d}
\end{eqnarray}
where the values of $R_1^1,S_3^3,P_{ai},$ are taken respectively
from (\ref {ricci1}), (\ref{ricci2}), (\ref{ricci3}),
(\ref{ricci4}).

By using the equations (\ref{einsteq3c}) and (\ref{einsteq3d}) we
can define the N--coefficients (\ref{ncoef}), $q_i(x^k,z)$ and
$n_i(x^k,z),$ if the functions $g_i(x^k)$ and $h_i(x^k,z)$ are
known as respective solutions of the equations (\ref{einsteq3a})
and (\ref{einsteq3b}).

Let consider an ansatz for a 4D d--metric of type
\begin{equation}
{\delta s}^2=g_1(x^k)(dx^1)^2+(dx^2)^2+h_3(x^i,t)(\delta
t)^2+h_4(x^i,t)(\delta y^4)^2, \label{dmetr4}
\end{equation}
where the $z$--parameter is considered to be the time like
coordinate and the energy momentum d--tensor is taken $$ \Upsilon
_\alpha^\beta = [p_1,p_2,-\varepsilon, p_4=p].$$ The aim of this
section is to analyze the system of partial differential equations
forllowing form the Einsteni field equations for these d--metric
and energy--momentum d--tensor.

\subsection{The h--equations}
The Einstein equations (\ref{einsteq3a}), with the Ricci h--tensor
(\ref {ricci1}), for the d--metric (\ref{dmetr4}) transform into
\begin{equation}
\label{hbh1}\frac{\partial ^2g_1}{\partial (x^1)^2}-\frac
1{2g_1}\left( \frac{\partial g_1}{\partial x^1}\right) ^2+2\kappa
\varepsilon g_1=0.
\end{equation}
By introducing the coordinates $\chi ^i=x^i/\sqrt{\kappa
\varepsilon}$  and the variable
\begin{equation}
q=g_1^{\prime }/g_1, \label{eq4c}
\end{equation}
where by 'prime' in this Section is considered the partial
derivative $\partial /\chi ^2,$ the equation (\ref{hbh1})
transforms into
\begin{equation}
\label{hbh2}q^{\prime }+\frac{q^2}2+2\epsilon =0,
\end{equation}
where the vacuum case should be parametrized for $\epsilon =0$
with $\chi ^i=x^i$ and $\epsilon =-1$ for a matter state with
$\varepsilon = - p$.

The integral curve of (\ref{hbh2}), intersecting a point $\left(
\chi _{(0)}^2,q_{(0)}\right) ,$ considered as a differential
equation on $\chi ^2$
is defined by the functions \cite{kamke}%
\begin{eqnarray}
q &=&
\frac{q_{(0)}}{1+\frac{q_{(0)}}2 %
\left( \chi ^2-\chi _{(0)}^2\right) },\qquad   \epsilon =0; \label{eq3a} \\%
 q & = & \frac{q_{(0)}-2\tan \left( \chi ^2-\chi _{(0)}^2\right) }%
 {1+\frac{q_{(0)}}2\tan \left( \chi ^2-\chi _{(0)}^2\right) },\qquad %
 \epsilon <0.   \label{eq5c} %
\end{eqnarray}

Because the function $q$ depends also parametrically on variable
$\chi ^1$ we can consider functions $\chi _{(0)}^2=\chi
_{(0)}^2\left( \chi ^1\right) $ and $q_{(0)}=q_{(0)}\left( \chi
^1\right) .$

We elucidate the nonvacuum case with $\epsilon <0.$ The general
formula for the non--trivial component of h--metric is to be
obtained after
integration on $\chi ^1$ of (\ref{eq4c}) by using the solution (\ref{eq5c})%
$$
 g_1\left( \chi ^1,\chi ^2\right) = g_{1(0)} \left( \chi ^1\right)
 \left\{ \sin [\chi ^2-\chi _{(0)}^2 \left( \chi ^1\right) ]+\arctan \frac 2{q_{(0)}\left(
\chi ^1\right) }\right\} ^2, $$ for $q_{(0)}\left( \chi ^1\right)
\neq 0,$ and
\begin{equation}
\label{btzlh3}g_1\left( \chi ^1,\chi ^2 \right) =g_{1(0)}\left(
\chi ^1\right) \ \cos ^2[\chi ^2-\chi _{(0)}^2\left( \chi
^1\right) ]
\end{equation}
for $q_{(0)}(\chi ^1) =0,$ where $g_{1(0)}(\chi^1), \chi
_{(0)}^2(\chi ^1) $ and $q_{(0)}(\chi^1) $ are some functions of
necessary smoothness class on variable $\chi ^1.$

For simplicity, in our further considerations we shall apply the solution (%
\ref{btzlh3}).

\subsection{The v--equations}
For the ansatz (\ref{dmetr4}) the Einstein equations
(\ref{einsteq3b}) with the Ricci h--tensor (\ref{ricci2})
transforms into
\begin{equation}
\label{heq}
\frac{\partial ^2h_4}{\partial t^2} - %
\frac 1{2h_4}\left( \frac{\partial h_4}{\partial t}\right) ^2
-\frac 1{2h_3}\left( \frac{\partial h_3}{\partial t}\right) \left(
\frac{\partial h_4}{\partial t}\right) -
\frac \kappa 2\Upsilon _1h_3h_4=0 \nonumber %
\end{equation}
(here we write down the partial derivatives on $t$ in explicit
form) which relates some first and second  order partial on $z$
derivatives of diagonal components $h_a(x^i,t)$ of a  v--metric
with a source $$\Upsilon_1(x^i,z)=\kappa \Upsilon  _1^1=\kappa
\Upsilon _2^2 = p_1=p_2$$ in the h--subspace. We can consider as
unknown  the function $h_3(x^i,t)$ (or, inversely, $h_4(x^i,t))$
for some compatible  values of $h_4(x^i,t)$ (or $h_3(x^i,t))$ and
source $\Upsilon_1(x^i,t).$    By introducing a new variable
$\beta =h_4^{*}/h_4$ the equation (\ref{heq})  transforms into
\begin{equation}  \label{heq1}\beta ^{*}+\frac 12\beta ^2-\frac{\beta
h_3^{*}}{2h_3}-2\kappa  \Upsilon _1h_3=0
\end{equation}
which relates two functions $\beta \left( x^i,t\right) $ and
$h_3\left(
x^i,t\right) .$ There are two possibilities: 1) to define $\beta $ (i. e. $%
h_4)$ when $\kappa \Upsilon _1$ and $h_3$ are prescribed and,
inversely 2) to find $h_3$ for given $\kappa \Upsilon _1$ and
$h_4$ (i. e. $\beta );$ in both cases one considers only ''*''
derivatives on $t$--variable with coordinates $x^i$ treated as
parameters.

\begin{enumerate}
\item  In the first case the explicit solutions of (\ref{heq1}) have to be
constructed by using the integral varieties of the general Riccati
equation
\cite{kamke} which by a corresponding redefinition of variables, $%
t\rightarrow t\left( \varsigma \right) $ and $\beta \left(
t\right) \rightarrow \eta \left( \varsigma \right) $ (for
simplicity, we omit dependencies on $x^i)$ could be written in the
canonical form $$ \frac{\partial \eta }{\partial \varsigma }+\eta
^2+\Psi \left( \varsigma \right) =0 $$ where $\Psi $ vanishes for
vacuum gravitational fields. In vacuum cases the Riccati equation
reduces to a Bernoulli equation which (we can use the former
variables) for $s(t)=\beta ^{-1}$ transforms into a linear
differential (on $t)$ equation,
\begin{equation}
\label{heq1a}s^{*}+\frac{h_3^{*}}{2h_3}s-\frac 12=0.
\end{equation}

\item  In the second (inverse) case when $h_3$ is to be found for some
prescribed $\kappa \Upsilon _1$ and $\beta $ the equation
(\ref{heq1}) is to be treated as a Bernoulli type equation,
\begin{equation}
\label{heq2}h_3^{*}=-\frac{4\kappa \Upsilon _1}\beta (h_3)^2+\left( \frac{%
2\beta ^{*}}\beta +\beta \right) h_3
\end{equation}
which can be solved by standard methods. In the vacuum case the squared on $%
h_3$ term vanishes and we obtain a linear differential (on $t)$
equation.
\end{enumerate}

Finally, in this Section we conclude that  the system of equations
(\ref{einsteq3b}) is
 satisfied by arbitrary functions
$$ \nonumber h_3=a_3(\chi ^i)\mbox{ and }h_4=a_4(\chi ^i). $$ If
v--metrics depending on three coordinates are introduced,
$h_a=h_a(\chi ^i,t),$ the v--components of the Einstein equations
transforms into (\ref {heq}) which reduces to (\ref{heq1}) for
prescribed values of $h_3(\chi ^i,t),\,$ and, inversely, to
(\ref{heq2}) if $h_4(\chi ^i,t)$ is prescribed.

\subsection{H--v equations}

For the ansatz (\ref{dmetr4}) with $h_4 = h_4 (x^i)$ and a
diagonal energy--momentum
 d--tensor the h--v--com\-po\-nents of Einstein equations (\ref{einsteq3c})
 and (\ref{einsteq3d}) are written respectively as
 \begin{equation} \label{einsteq5c}
 P_{5i}=\frac{q_i}{2h_3} [ {({\frac{\partial h_3}{\partial t} })}^2 -
 \frac{{\partial} ^2 h_3}{{\partial t}^2} ]=0,
 \end{equation}
 and
  \begin{equation} \label{einsteq5d}
 P_{6i}= {\frac{h_4}{4 (h_3)^2}}
 {\frac{\partial n_i}{\partial t}}{\frac{\partial h_3}{\partial t}} -
 {\frac{h_4}{2h_3}}{\frac{\partial ^2 n_i}{{\partial t}^2}} = 0.
 \end{equation}

The equations (\ref{einsteq5c}) are satisfied by arbitrary
coefficients
 $q_i(x^k,t)$ if the d--metric coefficient $h_3$ is a solution of
 \begin{equation} \label{ncomp}
 {({\frac{\partial h_3}{\partial t} })}^2 -  \frac{{\partial} ^2 h_3}{{\partial t}^2} = 0
 \end{equation}
 and the $q$--coefficients must vanish if this
 condition is not satisfied. In the last case we obtain a $3+1$ anisotropy.

 The general solution of equations (\ref{einsteq5d}) are written in the form
 $$n_i= l^{(0)}_i (x^k) \int \sqrt {|h_3 (x^k,t)|} dt + n^{(0)}_i (x^k)$$
 where $l^{(0)}_i (x^k)$ and $n^{(0)}_i (x^k)$ are arbitrary functions on $x^k$ which
 have to be defined by some boundary conditions.

\section{ Cos\-mo\-lo\-gi\-cal La--So\-lu\-ti\-ons}

The aim of this section is to construct two classes of solutions
of Einstein
 equations describing
 Friedman--Robertson--Walker (FRW) like universes with corresponding symmetries
  or rotational ellipsoid (ellongated and flattend) and torus.

\subsection{Rotation ellipsoid FRW universes}

We proof that there are  cosmological solutions constructed as
locally anisotropic
 deformations of the FRW spherical symmetric solution to the rotation ellipsoid configuration.
 There are two types of rotation ellipsoids, elongated and flattened ones. We
examine both cases  of such horizon configurations.

\subsubsection{Rotation elongated ellipsoid configuration}

An elongated rotation ellipsoid hypersurface is given by the
formula \cite {korn}
\begin{equation}
\label{relhor}\frac{{x}^2+{y}^2}{\sigma ^2-1}+\frac{%
{z}^2}{\sigma ^2}={\rho }^2,
\end{equation}
where $\sigma \geq 1,$ $x,y,z$ are Cartezian coordinates and
${\rho }$ is similar to the radial coordinate in the spherical
symmetric case.

The 3D special coordinate system is defined
\begin{eqnarray}
{x} &=&{\rho}\sinh u\sin v\cos \varphi ,\ {y}={\rho}\sinh u\sin
v\sin \varphi ,\nonumber \\ {z}&=& {\rho}\ \cosh u \cos v,
\nonumber
\end{eqnarray}
where $\sigma =\cosh u,(0\leq u<\infty ,\ 0\leq v\leq \pi ,\ 0\leq
\varphi <2\pi ). $\ The hypersurface metric (\ref{relhor}) is
\begin{eqnarray}
g_{uu} &=& g_{vv}={\rho}^2\left( \sinh ^2u+\sin ^2v\right) ,
 \label{hsuf1} \\
g_{\varphi \varphi } &=&{\rho}^2\sinh ^2u\sin ^2v.
 \nonumber
\end{eqnarray}

Let us introduce a d--metric of class (\ref{dmetr4})
\begin{equation}
\label{rel1}\delta s^2 = g_1(u,v)du^2+dv^2 +  h_3\left( u,v,\tau
\right) \left( \delta \tau \right) ^2+ h_4\left( u,v\right) \left(
\delta\varphi \right) ^2,
\end{equation}
where $x^1=u, x^2=v,$  $y^4= \varphi,$ $y^3=\tau$ is the time like
cosmological coordinate and $\delta \tau$ and $\delta \varphi $
are N--elongated differentials.

As a particular solution of (\ref{rel1}) for the h--metric we
choose (see (\ref{btzlh3})) the coefficient
\begin{equation}
\label{relh1h}g_1(u,v)=\cos ^2v
\end{equation}
and set for the v--metric components
\begin{equation}
\label{relh1}h_3(u,v,\tau)=-\frac 1{\rho ^2(\tau) (\sinh ^2u+\sin
^2v)}
\end{equation}
and
\begin{equation}
h_4(u,v,\tau)=\frac {\sinh ^2u \sin ^2v}{(\sinh ^2u+\sin ^2v)}.
\label{relh2}
\end{equation}

The set of coefficients (\ref{relh1h}),(\ref{relh1}), and
(\ref{relh2}), for the d--metric (\ref{rel1}, and of $q_i=0$ and
$n_i$ being solutions of (\ref{ncomp}), for the N--connection,
defines a solution of the Einstein equations (\ref{einsteq2}).

The physical treatment of the obtained solutions follows from the
locally isotropic limit of a conformal transform of this
d--metric: Multiplying (\ref{rel1}) on $$ {\rho ^2(\tau) (\sinh
^2u+\sin ^2v)}, $$ and considering $cos ^2v \simeq 1$ and $n_i
\simeq = 0$ for locally isotropic spacetimes we get the interval
\begin{eqnarray}
ds^2 &= &- d \tau ^2 + \rho ^2 (\tau) [(\sinh ^2u+\sin ^2v)(du^2 +
dv^2) +
 {\sinh ^2u} {\sin ^2v} d\varphi ^2] \nonumber \\
{ }&{ }&  \mbox{for ellipsoidal coordinates on hypersurface
(\ref{hsuf1})}; \nonumber \\ { } &=& - d \tau ^2 + \rho ^2 (\tau)
[dx^2 + dy^2 + dz^2]\  \mbox{for Cartezian coordinates}, \nonumber
\end{eqnarray}
which defines just the Robertson--Walker metric.

So, the d--metric (\ref{rel1}), the coefficients of N--connection
being solutions of (\ref{einsteq3c}) and (\ref{einsteq3d}),
describes a 4D cosmological solution of the Einstein equations
when instead of a spherical symmetry
 one has a locally anisotropic deformation to the symmetry
of rotation elongated ellipsoid. The explicit dependence on time
$\tau$ of the
 cosmological factor $\rho$ must be constructed by using additionally the matter
 state equations for a cosmological model with local anisotropy.

\subsubsection{Flattened rotation ellipsoid coordinates}

In a similar fashion we can construct a locally anisotropic
deformation of the  FRW metric with the symmetry of flattened
rotation ellipsoid.
 The parametric equation for a such hypersurface is \cite{korn}
$$ \frac{{x}^2+{y}^2}{1+\sigma ^2}+\frac{{z}^2}{\sigma ^2}={\rho
}^2, $$ where $\sigma \geq 0$ and $\sigma =\sinh u.$

The proper for ellipsoid 3D  space coordinate system is defined%
\begin{eqnarray}
{x} &=&{\rho}\cosh u\sin v\cos \varphi , y = {\rho} \cosh u\sin v
\sin \varphi \nonumber \\ {z} &=& {\rho} \sinh u\cos v, \nonumber
\end{eqnarray}
where $0\leq u<\infty ,\ 0\leq v\leq \pi ,\ 0\leq \varphi <2\pi .$

The hypersurface metric is
\begin{eqnarray}
g_{uu} &=& g_{vv}={\rho}^2\left( \sinh ^2u+\cos ^2v\right) ,
 \nonumber \\
g_{\varphi \varphi } &=&{\rho}^2\sinh ^2u\cos ^2v.
 \nonumber
\end{eqnarray}
In the rest the cosmological la--solution is described by the same
formulas as in the previous subsection but with respect to new
canonical coordinates for flattened rotation ellipsoid.

\subsection{Toroidal FRW universes}

Let us construct a cosmological solution of the Einstein equations
with toroidal
symmetry. The hypersurface formula of a torus is \cite{korn}%
$$ \left( \sqrt{{x}^2+{y}^2}-{\rho }\ c\tanh \sigma \right)
^2+{z}^2= \frac{{\rho }^2}{\sinh ^2\sigma }. $$ The 3D space
coordinate system is defined
\begin{eqnarray}
{x} &=& \frac{{\rho}\sinh \alpha \cos \varphi }{\cosh \alpha
-\cos\sigma },
 \qquad
{y} = \frac{{\rho}\sin \sigma \sin \varphi }{\cosh \alpha -\cos
\sigma }, \nonumber \\ {z} &=& \frac{{\rho}\sinh \sigma }{\cosh
\tau -\cos \sigma }, \nonumber \\ &{}& \left( -\pi <\sigma <\pi ,
0\leq \alpha <\infty ,0\leq \varphi <2\pi \right) .
 \nonumber
\end{eqnarray}
The hypersurface metric is
\begin{equation}
\label{mtor}g_{\sigma \sigma }=g_{\alpha \alpha }=\frac{{\rho
}^2}{\left( \cosh \alpha -\cos \sigma \right) ^2}, g_{\varphi
\varphi }=\frac{{\rho }^2\sin ^2\sigma } {\left( \cosh \alpha
-\cos \sigma \right) ^2}.
\end{equation}

The d--metric of class (\ref{dmetr4}) is chosen
\begin{equation}
\label{mtora}\delta s^2 = g_1(\alpha)d\sigma ^2+d \alpha ^2 +
 h_3\left(\sigma, \alpha, \tau \right) \left( \delta \tau \right) ^2+
h_4\left( \sigma \right) \left( \delta\varphi \right) ^2,
\end{equation}
where $x^1= \sigma , x^2= \alpha ,$  $y^4= \varphi,$ $y^3=\tau$ is
the time like cosmological coordinate and $\delta \tau$ and
$\delta \varphi $ are N--elongated differentials.

As a particular solution of (\ref{mtor}) for the h--metric we
choose (see (\ref{btzlh3})) the coefficient
\begin{equation}
\label{relh2h}g_1(\alpha)=\cos ^2\alpha
\end{equation}
and set for the v--metric components
\begin{eqnarray}
h_3(\sigma, \alpha,\tau) &=&-\frac {{(\cosh\alpha - \cos \sigma
)}^2 }{\rho ^2(\tau) }
 \nonumber \\
h_4(\sigma) & = &\sin ^2 \sigma. \label{relh2a}
\end{eqnarray}

Multiplying (\ref{mtora}) on $$ \frac{\rho ^2(\tau)}{{(\cosh
\alpha - \cos \sigma)}^2}, $$ and considering $cos \alpha \simeq
1$ and $n_i \simeq = 0$ in the locally isotropic limit we get the
interval $$ ds^2 = - d \tau ^2 + \frac {\rho ^2
(\tau)}{{(\cosh\alpha - \cos \sigma )}^2 } [(d \sigma ^2 + d\alpha
^2 +
  {\sin ^2 \sigma } d\varphi ^2]
  $$
  where the space part is just the torus hypersurface metric (\ref{mtor}).

So, the set of coefficients (\ref{relh2h}) and (\ref{relh2a}), for
the d--metric (\ref{mtora}, and of $q_i=0$ and $n_i$ being
solutions of (\ref{ncomp}), for the N--connection, defines a
cosmologica solution of the Einstein equations (\ref{einsteq2})
with the torus symmetry,
 when the explicit form of the function $\rho (\tau)$ is to be defined by considering some additional equations
  for the matter state (for instance, with a scalar field defining  the torus inflation).

\section{Outlook and Concluding Remarks}

In this paper we have developed the method of anholonomic frames
on (pseudo) Riemannian spacetimes by considering associated
nonlinear connection (N--con\-nec\-ti\-on) strucutres. We provided
a rigorous geometric background for description of gravitational
systems with mixed
 holonomic and anholonomic (anisotropic) degrees of freedom by considering first and
 higher order anisotropies induced by anholonomic constraints and corresponding frame
 bases.

 The first key result of this paper is the proof that generic anisotropic structures of different
  order are contained in the Einstein theory.  We reformulated the tensor and linear connection
   formalism for (pseudo) Riemannian spaces enables with N--connections and computed the
   horizonal--vertical splitting, with respect to anholonomic frames with associated N--connections,
   of the Einstein equations. The (pseudo) Riemannian spaces enabled with  compatible
   anholonomic frame and associated N--connection structures and the metric being a solution of
    the Einstein equations were called as locally anisotropic spacetimes (la--spacetimes).

   The next step was the definition of gauge field interactions on la--spacetimes. We have applied the
    bundle formalism and extended it to the case of bases being la--spacetimes  and considered a ´pure´
     geometric method of deriving the Yang--Mills equations for generic
      locally anisotropic gauge interactions, by genalizing the absolut differential calculus
       and dual forms symmetries for la--spacetimes.

   The second key result was the proof by geometric methods that the Yang--Mills equations for
   a correspondingly defined Cartan
    connection in the bundle of affine frames on la--spacetimes are equivalent to the Einstein  equations
     with anholonomic (N--connection) structures (the original Popov--Dikhin papers \cite{p,pd}
      were for the locally isotropic spaces). The result was obtained by applying an auxiliary
      bilinear form on the tipical fiber because of degeneration of the Killing form for the affine
       groups. After projection on base spacetimes the dependence on auxiliar values is elliminated.

       We analyzed  also a variant of variational gauge locally anisotropic gauge theory by considering
    a minimal extension of the affine  structural group to the de Sitter one, with a nonlinear
     realization for the gauge group as one was performed in a locally isotropic version in Tseytlin's
      paper \cite{tseyt}. If some former our works \cite{vg,vhp} where devoted to extensions of
       some models of gauge gravity to generalized Lagrange and Finsler spaces, in this paper
       we demonstrated which manner we could manage with anisotropies arrising in locally isotropic,
       but with anholonomic structures, variants of gauge gravity. Here it should be emphasized
       that anisotropies of different type (Finsler like, or more general ones) could be induced
       in all variants of gravity theories dealing with frame (tetrad, vierbiend, in four dimensions)
       fields and decompisitions of geometrical and physical objects in comonents with respect
        to such frames and associated N--connections.  In a similar fashion anisotropies could arise
    under nontrivial reductions from higher to lower dimensions in Kaluza--Klein theories;
    in this case the N--connection should be treated as a  splitting field
    modelling the anholonomic (anisotropic) character of some degrees of  freedom.

    The third basic result is the construction of a new class of solutions, with generic
    local anisotropy, of the Einstein equations. For simplicity, we defined
    these solutions in the framework of general relativity, but they can be
    removed to various variants of gauge and spinor gravity by using corresponding
    decompositions of the metric into the frame fields. We note that the obtained class of
    solutions also holds true for the gauge models of gravity which, in this paper,
    were constructed to be equivalent to the Einstein theory.

    In explicit form we considered the metric ansatz
$$ ds^2=g_{\alpha \beta }\ du^\alpha du^\beta $$ when  $g_{\alpha
\beta}$  are parametrized by  matrices of type
\begin{equation}
\left[
\begin{array}{cccc}
g_1+q_1{}^2h_3+n_1{}^2h_4 & 0 & q_1h_3 & n_1h_4 \\ 0 &
g_2+q_2{}^2h_3+n_2{}^2h_4 & q_2h_3 & n_2h_4 \\ q_1h_3 & q_2h_3 &
h_3 & 0 \\ n_1h_4 & n_2h_4 & 0 & h_4
\end{array}
\right]   \label{ansatz2}
\end{equation}
with coefficients being some functions of necessary smo\-oth class $%
g_i=g_i(x^j),q_i=q_i(x^j,t),n_i=n_i(x^j,t),$ $h_a=h_a(x^j,t).$
Latin indices run respectively $i,j,k,...$ $=1,2$ and
$a,b,c,...=3,4$ and the local coordinates are denoted $u^\alpha
=(x^i,y^3=t,y^4),$ where $t$ is treated as a timelike coordinate.
A metric  (\ref{ansatz2})  can be diagonalized,
\begin{equation} \label{diag}
\delta s^2=g_i(x^j)\left( dx^i\right) ^2+ h_a(x^j,t)\left( \delta
y^a\right)^2,
\end{equation}
with respect to anholonomic frames (\ref{dder}) and (\ref{ddif}),
here we write down only the 'elongated' differentials $$ \delta
t=dz+q_i(x^j,t)dx^i,\ \delta y^4=dy^4+n_i(x^j,t)dx^i. $$ The
ansatz (\ref{ansatz2}) was formally introduced  in \cite{v2} in
order to construct locally anisotropic black hole solutions; in
this paper we applied it to cosmological la--spacetimes. In
result, we get new metrics which describe locally
 anisotropic  Friedman--Robertson--Walker like universes  with the spherical symmetry de\-form\-ed
  to that of rotation (ellongated and/or flattened) ellipsoid and torus. Such solutions
  are contained in  general relativity: in the symplest diagonal form they are parametrized
  by distinguished metrics of type (\ref{diag}), given with respect to anholonomic bases, but
  could be also  described equivalently with respect to a coordinate base by matrices of
  type (\ref{ansatz2}). The topic of construction of cosmological models with generic
   spacetime and matter field distribution and fluctuation anisotropies is under consideration.

  Now, we point the item of definition of reference frames in gravity theories:
 The form of basic field equations and fundamental laws in general relativity do not depend
 on choosing of coordinate systems and frame bases. Nevetheless, the problem of fixing
 of an adequate system of reference is also a very important physical task which is not solved
 by any dynamical equations but following some arguments on measuring of phyical observables,
  imposed symetry of interactions, types of horizons and singularities, and by taken into consideration
   the posed Cauchy problem. Having fixed a class of frame variables, the frame coefficients
   being presented in the Einstein equations, the type of constructed solution depends on the
    chosen holonomic or anholonomic frame structure. As a result one could model various forms
    of anisotropies in the framework of the Einsten theory (roughly, on  (pseudo) Riemannian
    spacetimes with  corresponding anholonomic frame structures it is possible to model
    Finsler like metrics, or more general ones with anisotropies). Finally, it should be noted that
    such questions on stability of obtained solutions, analysis of energy--momentum conditions
    should be performed in the simplest form with respect to the chosen class of anholonomic frames.

\vskip10pt {\bf Acknoweledgments} \vskip5pt The S. V. work was
supported by the
 German Academic Exchange Service (DAAD).


\end{document}